\documentclass[aps,twocolumn]{revtex4-2}
\usepackage{amsmath,amssymb}
\usepackage{slashed}
\usepackage{graphicx}
\usepackage{bm}
\usepackage[T1]{fontenc}
\usepackage[utf8]{inputenc}
\usepackage{gauss} 
\usepackage{tikz}
\usepackage{ytableau}
\usepackage[top=1.0in,bottom=1.0in,left=1.0in,right=1.0in]{geometry}
\usepackage[colorlinks,linkcolor=blue,citecolor=blue]{hyperref}
\usepackage{comment}
\usepackage{ytableau}

\newcommand{\YTtwoone}{%
\begin{array}{|c|c|}\hline
\mbox{}&\mbox{}\\ \hline
\end{array}
\kern-0.4pt
\raisebox{-1.75ex}{\begin{array}{|c|}\hline
\mbox{}\\ \hline
\end{array}}}
\newcommand{\YTrho}{%
\begin{array}{|c|c|}\hline
1&2\\ \hline
\end{array}
\kern-0.4pt
\raisebox{-1.75ex}{\begin{array}{|c|}\hline
3\\ \hline
\end{array}}}
\newcommand{\YTlambda}{%
\begin{array}{|c|c|}\hline
1&3\\ \hline
\end{array}
\kern-0.4pt
\raisebox{-1.75ex}{\begin{array}{|c|}\hline
2\\ \hline
\end{array}}}
\newcommand{\be}{\begin{equation}}
\newcommand{\ee}{\end{equation}}
\newcommand{\bea}{\begin{eqnarray}}
\newcommand{\eea}{\end{eqnarray}}
\newcommand{\ba}{\begin{eqnarray}}
\newcommand{\ea}{\end{eqnarray}}

\begin{document}

\title{Hybrid hadrons at rest and on the light front}

\author{Edward Shuryak}
\email{edward.shuryak@stonybrook.edu}
\affiliation{Center for Nuclear Theory, Department of Physics and Astronomy, Stony Brook University, Stony Brook, New York 11794-3800, USA}

\author{Ismail Zahed}
\email{ismail.zahed@stonybrook.edu}
\affiliation{Center for Nuclear Theory, Department of Physics and Astronomy, Stony Brook University, Stony Brook, New York 11794-3800, USA}

\begin{abstract}
We present  a unified description of heavy hybrid hadrons based on a constituent-gluon picture embedded in the Born-Oppenheimer (BO) framework. In this approach, the gluonic excitation is treated as a dynamical quasiparticle with a mass generated by instanton-induced interactions. We propose a simple variational derivation of the BO potentials.
The main focus of the paper is the derivation of light-front wave functions for hybrid systems, specifically for the $ccg$ and $qqqg$ cases. We employ both variational methods and numerical solutions of the Schr\"odinger equation in momentum representation. Using the resulting wave functions, we compute the gluon PDFs for these systems.
\end{abstract}


\maketitle

\section{Introduction}
\subsection{Brief overview}

A recent active period in hadronic spectroscopy started about two decades ago with multiple discoveries of hadronic states that are neither mesons nor baryons. First, in charmonium spectroscopy, these were the XYZ resonances, which clearly do not fit a $\bar c c$ interpretation. Those with nonzero isospin are naturally interpreted as tetraquarks $\bar c c \bar q q$, and are now commonly denoted as $T_{c\bar c}$ states. Others, specifically in the vector $J^{PC}=1^{--}$ channel, are still called $\psi$, retaining the same names as   charmonium states.

Further studies of these ``non-charmonium" states revived the long-standing issue of so-called hybrid states, with structure $\bar Q g Q$, involving an additional ``constituent gluon". Those are the subject of our work.

Before going into technical details, let us discuss  simple qualitative estimates for the masses of tetraquarks and hybrid states. We begin by proposing a crude relation between meson and glueball masses. The constituent quark mass is about $M_q \approx 350\,\text{MeV}$, while the constituent gluon mass, fitted from glueball spectra \cite{Shuryak:2026grt,Shuryak:2026pqt}, is $M_g \approx 900\,\text{MeV}$. Their ratio ($\approx 2.57$) is close to the Casimir scaling ratio $9/4$, which governs the strength of perturbative Coulomb interactions. The adjoint confining potential is also stronger by approximately the same factor \cite{Shuryak:2026grt,Shuryak:2026pqt}. This suggests that Casimir scaling may be approximately used for  glueball-to-meson mass ratios as well.

Proceeding further, the mass difference between hybrids and tetraquarks should include the difference between effective gluon mass and twice the effective quark mass. Assuming Casimir scaling, one estimates
\[
M_g - 2M_q \approx M_g/4 \approx 225\,\text{MeV}.
\]
This combination should therefore appear in the difference between hybrid and tetraquark masses:
\be 
M(\text{hybrids}) - M(\text{tetras}) \approx M_g - 2M_q \sim 200\,\text{MeV}.
\label{eqn_naive_diff}
\ee
Of course, interaction terms in the Hamiltonian, depending on quantum numbers, can contribute corrections of comparable size, but experience suggests partial cancellations. Thus, in a statistical sense, one expects hybrids to be {\em slightly heavier than tetraquarks}, by about $200\,\text{MeV}$.

Let us now compare these naive estimates with  sophisticated theoretical results based on the Born-Oppenheimer approximation \cite{Berwein:2024ztx} for charmonium-based tetraquarks and hybrids. Rather than comparing the models directly, we consider their interpretation of experimentally observed states (see ``conclusions" and Table XII of that work). The lowest tetraquarks for five $J^{PC}$ channels are   (masses in MeV)
$$
1^{--} \, \psi(4230),\psi(4360), \quad 1^{++} \, \chi_{c1}(3872),
$$
$$
0^{++}\, \chi_{c0}(3915), \quad 0^{-+} \, \chi_{c0}(3940), \quad
0^{-} \, T_{cc0}(4240),
$$
while the lowest exotic hybrids are projected to be
$$
1^{--} \,\psi(4360),\psi(4390), \quad
1^{-+} \,X(4630).
$$
Although it is too early for a channel-by-channel comparison, one observes that hybrids are indeed systematically a bit heavier than tetraquarks.

The importance of this (perhaps overly long) estimates lies in supporting the assumption of a particular value of the  gluon effective mass $M_g \sim 0.9\,\text{GeV}$. This conjecture is crucial for bridging the gap between glueball and hybrid spectroscopy and the perturbative evolution of PDFs, which we aim to address in this work.

Let us now briefly review the historical development of quarkonium-based tetraquarks and hybrids within the Born-Oppenheimer framework. While this approximation has been a cornerstone of molecular physics for about a century, it was introduced into hadronic spectroscopy by Juge, Kuti, and Morningstar \cite{Juge:1999ie}. Solving the Schr\"odinger equation with static potentials $V_{BO}(r)$, they demonstrated good agreement with direct (quenched) lattice simulations. Subsequent work \cite{Braaten:2014qka} further developed this approach, clarifying the structure of the Schr\"odinger equations and the associated quantum numbers.

Two main approximations are involved. The first is the {\em adiabatic approximation}, which uses static potentials and neglects relativistic corrections of order $O(v^2)$. This is better justified for heavier quarks and reduces the problem to a multichannel Schr\"odinger equation for the relative motion of $Q$ and $\bar Q$. The second approximation consists in separating channels with different quantum numbers $\Gamma$, ignoring possible mixing between them. Unlike the first, this approximation does not rely on the heavy-quark limit and requires further justification.

Let us now discuss how hybrid hadrons differ qualitatively from conventional ones. The key distinction lies in their color wave functions. In $\bar c c$ states, quarks form a color singlet $3 \otimes \bar 3 = 1$, whereas in hybrids they are in a color octet $3 \otimes \bar 3 = 8$, required to combine with a gluon into an overall singlet. Consequently, the effective potentials are entirely different: e.g. the Coulomb-like $1/r$ interaction is attractive in the singlet channel but repulsive in the octet channel.

In the absence of light quarks (and hence tetraquark channels), conventional quarkonia are described by a string potential
\be 
V_\Sigma(\text{large}, r)=\sigma r \left(1-\frac{\pi}{6 \sigma r}\right)^{1/2},
\ee
\be 
V_\Sigma(\text{small}, r)=-\frac{4}{3}\frac{\alpha_s(1/r)}{r}.
\ee
The total potential includes an additive constant $2M_Q+V_0$, with $V_0$ independent of the quark mass. Lattice calculations (quenched and unquenched) are well fitted by the Cornell form.

Hybrid potentials correspond to excited string configurations at large $r$:
\be 
V_{\Sigma^-_u}(\text{large}, r)=\sigma r \left(1-\frac{\pi (12 n_\Gamma-1)}{6 \sigma r}\right)^{1/2}.
\ee
At small distances, the interaction becomes a repulsive Coulomb potential:
\be 
V_\Sigma(\text{small}, r)=+\frac{1}{6}\frac{\alpha_s(1/r)}{r}+E_\Gamma.
\ee
In the limit $r \to 0$, one has a pointlike color-octet $\bar Q Q$ state bound with a gluon, known as a ``gluelump", which possesses additional symmetries.

The centrifugal term is modified to
$$
\left(L(L+1)-2\Lambda^2+\frac{J_\Gamma(J_\Gamma+1)}{M_Q r^2}\right).
$$

One possible direction for further studies is to follow the approach used for nucleons, where ``unquenching" is achieved by including effective gluons, as well as $\bar q q$ pairs in the form of $\pi$ or $\sigma$ mesons. These mechanisms should eventually explain the emergence of orbital motion and flavor asymmetries in antiquark PDFs, see e.g. \cite{Miesch:2025ael,He:2025dik}, and the precise
form of renormalization group for Hamiltonians used.

\subsection{Motivations and structure of the paper}

There are two main motivations for this work.

First, as newcomers to this area of hadronic spectroscopy, we focused on Born-Oppenheimer (BO) potentials. While these have been computed on the lattice since the pioneering studies, we find it desirable to derive them analytically, for example variationally, using simple Hamiltonians and ansatz wave functions. Although one does not expect high precision, the results turn out to be surprisingly close to lattice calculations (see Fig.~\ref{fig_BO}). We believe this derivation merits independent presentation; details are given in section \ref{sec_var}.

Solving  Schr\"odinger equation for $\Sigma_g^-$ hybrid
is straightforward and we do so in section \ref{sec_overlaps}. 
Similarity between BO potential for this channel and for
usual quarkonia $\Sigma_g^+$ (except small distances) leads
to similarity of spectra and the wave functions, see Fig.\ref{fig_WFs_Sigma_plus_minus}. As a consequence, the
overlap matrix is found to be close to $\delta_{mn}$. 
We discuss consequences of this observation for decays of
two lowest hybrid candidates, $\psi(4360)$ and $\psi(4660)$.
We find that the ratio of their observed widths does indeed
agree with such crude model of decay overlaps. 

The second motivation is the main one: we try to develop a {\em light-front (LF) description of tetraquarks and hybrids}. Our broader goal is to ``bridge hadronic spectroscopy with partonic observables", see \cite{Shuryak:2026pqt} for a review. Understanding hybrids on the light front, including their mixing with light-quark states, is an essential step toward realistic ``unquenched" wave functions that include both quark-antiquark pairs and gluons. By systematically increasing Fock components, one may ultimately replace perturbative PDF evolution with renormalization-group evolution of wave functions, from which all partonic observables can be derived.

We derive LF wave functions for two systems. The first is the quarkonium hybrids $\bar c c g$, discussed in section \ref{sec_ccg_LF}. Since the masses of charm quark and constituent gluon  are comparable, there is no clear separation of scales or $BO$ approximation, thus the problem is treated as a genuine three-body system. As in baryons, the forward wave function is defined on an equilateral triangle. We obtain it both variationally and by numerical solution of the Schr\"odinger equation in momentum space, and use it to compute the gluon PDF.

The second system is hybrid light-quark baryons, the four-body system $qqqg$,discussed in section \ref{sec_qqqg}. In this case the roles are reversed, since the constituent gluon is heavier than the quarks. We conclude by computing the gluon PDF for this system.

The paper is summarized in section \ref{sec_summary}, where we also briefly discuss outlook for further works.


\section{Variational derivation of the BO hybrid potential} \label{sec_var}

\begin{figure}[h]
    \centering
    \includegraphics[width=0.65\linewidth]{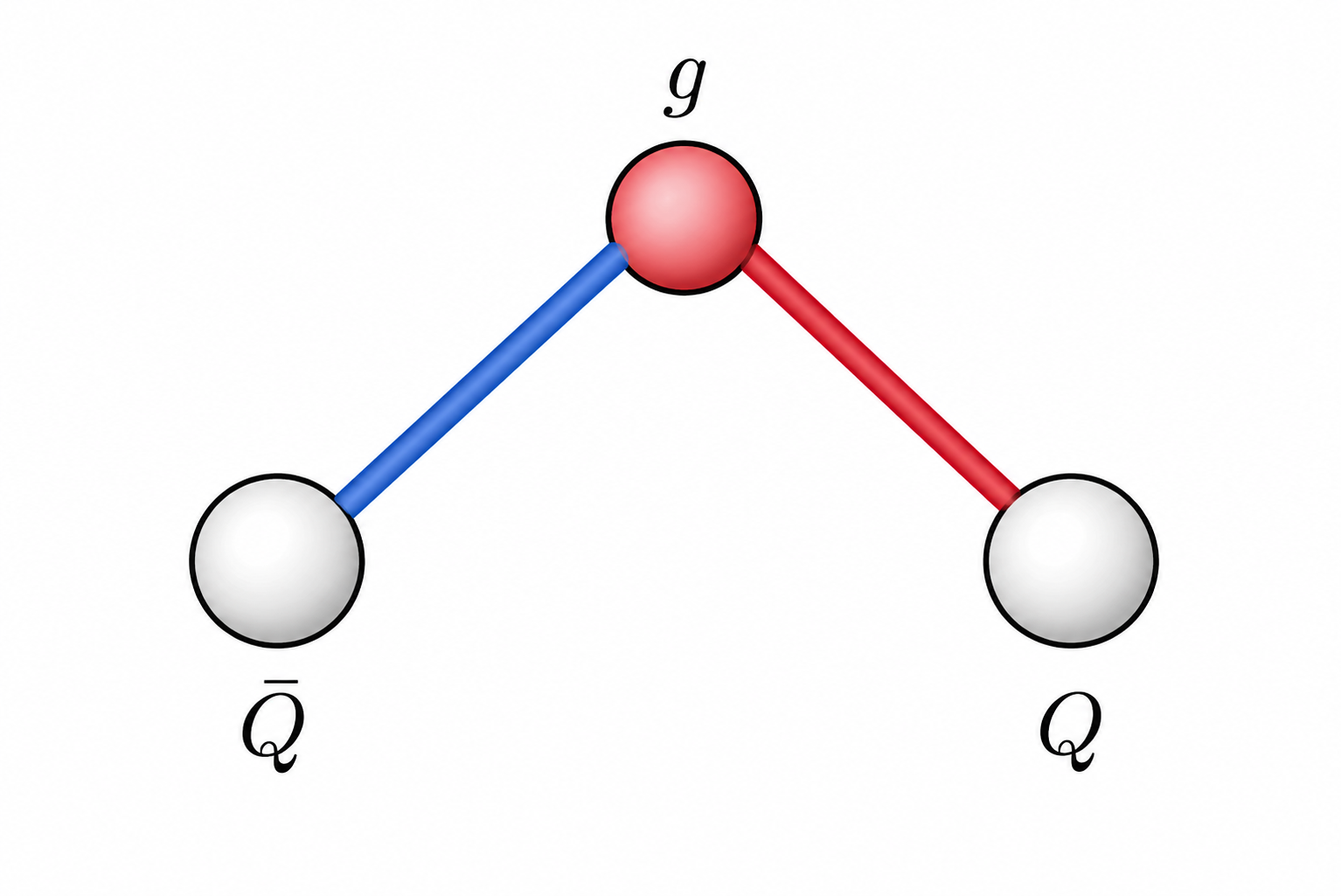}
    \caption{Schematic setting of the variational calculation: heavy quarks $\bar Q$ and $Q$ are static, and a ``constituent gluon" $g$ dynamical, connected to both by a linear potential.}
    \label{fig_BO_schematic}
\end{figure}

Schematic setting of the calculation
is shown in Fig.\ref{fig_BO_schematic}.
The heavy quarks are treated as static sources at fixed separation $r$, while the gluon $g$ is dynamical, interacting with
quarks by confining potentials.

\begin{figure}[b]
    \centering      \includegraphics[width=0.65\linewidth]{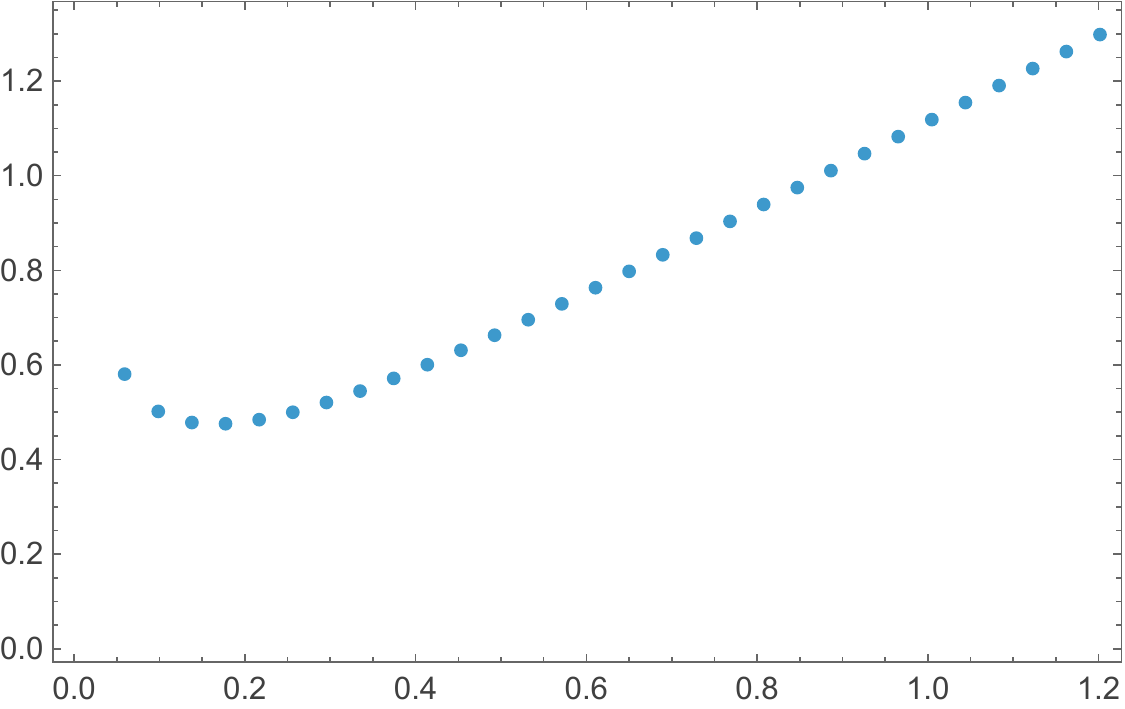}
    \includegraphics[width=0.85\linewidth]{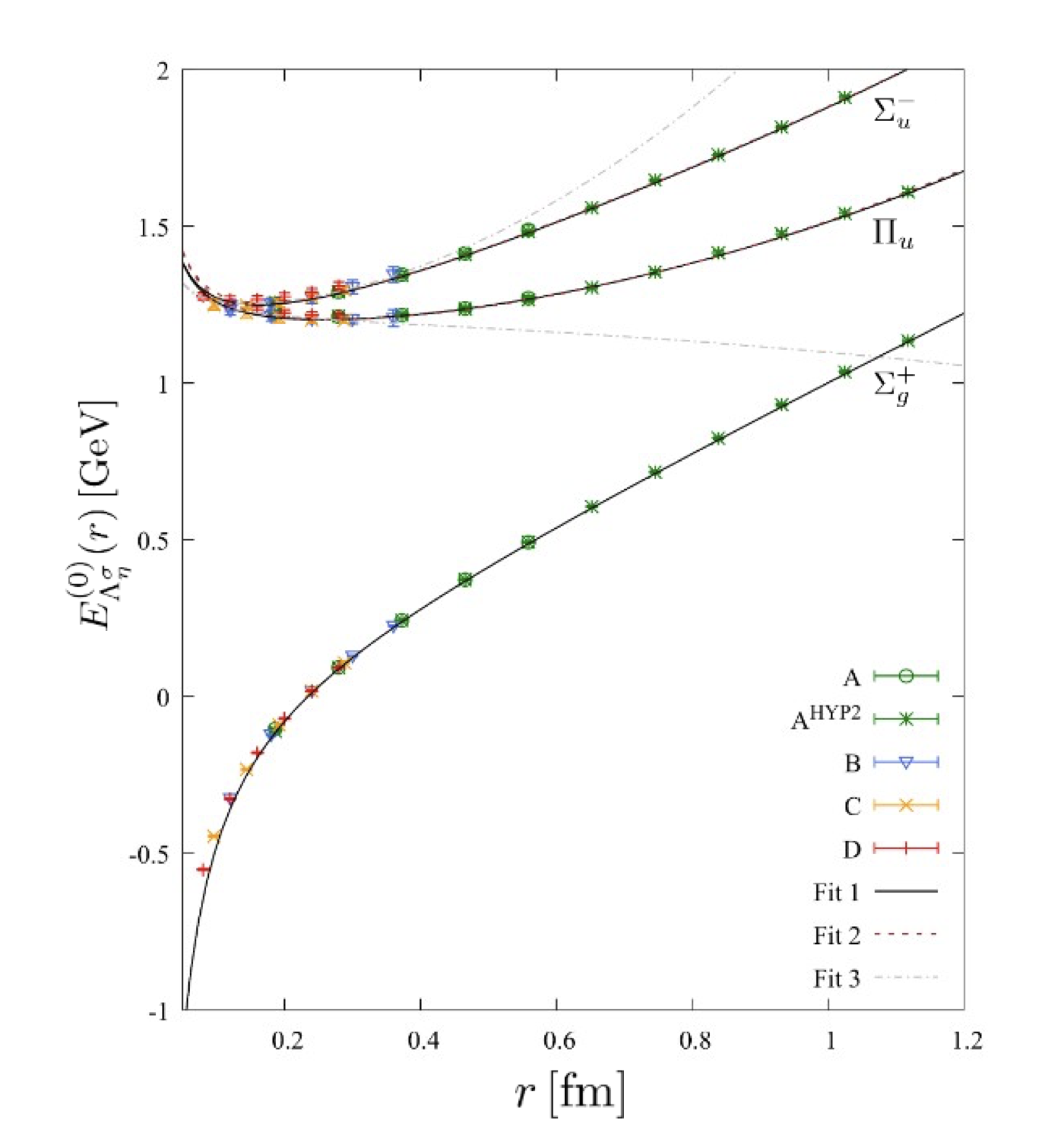}
    \caption{ (Upper) Result of our variational calculation of $\Sigma^-_u(r)$ potential versus quark-antiquark separation $r,  (fm)$.(Lower) The lowest hybrid static BO energies $\Pi_u$ and $\Sigma^-_u$ and the quarkonium static energy $\Sigma^+_g$, from lattice measurements reported in \cite{PhysRevD.105.054503}.\\
     }
    \label{fig_BO}
\end{figure}

The gluon wavefunction we approximate by a variational Gaussian ansatz, centered near the midpoint of the
heavy pair.  Two $Qg$ potentials are approximated by linear confining potentials.
This setting makes the molecular character of the hybrid explicit, and yields
closed expressions for the BO (adiabatic) potentials $V_\Gamma(r)$. 
The variational potential $\Sigma^-_u(r)$
we obtained is shown in Fig.\ref{fig_BO}(upper).
Modern versions of this potentials from lattice studies
are shown in Fig.\ref{fig_BO}(lower).

 Now we present details of the calculation. The heavy quark and antiquark are at fixed separation $r$
\begin{equation}
\vec R_Q = +\frac{\vec r}{2},
\qquad
\vec R_{\bar Q} = -\frac{\vec r}{2},
\qquad r = |\vec r| ,
\end{equation}
and let $\vec x$ denote the constituent-gluon coordinate measured from the
midpoint. The light-sector Hamiltonian at fixed $r$ is of the form
\ba
H_g(r)
&=&
\sqrt{\vec p^{\,2}+m_g^2}
+
V_{Qg}\!\left(\left|\vec x-\frac{\vec r}{2}\right|\right) \\
&+&
V_{\bar Qg}\!\left(\left|\vec x+\frac{\vec r}{2}\right|\right)
+
V_{\rm inst}(\vec x) +{\alpha_s(r) \over 6 r} \nonumber
\label{Hgfixedr}
\ea
with $m_g$ the dynamical constituent gluon mass. The short-distance part of the
quark-gluon interaction is attractive in the triplet channel and is modeled by
a Coulomb term,
\begin{equation}
V_{Qg}(\rho)=V_{\bar Qg}(\rho)=-\frac{\kappa}{\rho},
\qquad
\kappa = C_{Qg}\,\alpha_s ,
\end{equation}
with $C_{Qg}=\frac 32$, while the confining part is represented by  two string joining the gluon to the heavy sources,
\begin{equation}
V_{\rm conf}(\vec x;r)=\sigma\left(
\left|\vec x-\frac{\vec r}{2}\right|
+
\left|\vec x+\frac{\vec r}{2}\right|
\right).
\label{VY}
\end{equation}
At $r=0$ it reduces to a gluelump Hamiltonian with a constituent gluon moving in
the field of a compact octet source; at large $r$ it tends toward a string-like
configuration.

For the lowest gluonic excitation, corresponding to the $\Gamma=\Pi_u/\Sigma_u^-$
multiplet (with the spectroscopic notations briefly recalled  in the Appendix), the light field carries one unit of angular momentum about the
molecular axis. Choosing the internuclear axis along $z$, a convenient Gaussian
ansatz is
\begin{equation}
\psi_{1m}(\vec x)
=
{\cal N}_1\, x\, e^{-\beta^2 x^2/2}\, Y_{1m}(\hat x),
\qquad m=0,\pm1,
\label{gaussP}
\end{equation}
with variational width $\beta$. The $m=0$ component connects to the
$\Sigma_u^-$ channel, while $m=\pm1$ span the $\Pi_u$ doublet. In the
strict $r\to 0$ limit rotational symmetry is restored and these states become
degenerate, as expected for the lowest $J_g^{PC}=1^{+-}$ gluelump. For the
lowest $\Sigma_g^+$ channel one instead uses the $S$-wave ansatz
\begin{equation}
\psi_{00}(\vec x)
=
{\cal N}_0\, e^{-\beta^2 x^2/2} .
\label{gaussS}
\end{equation}
The variational energy is
\begin{equation}
V_\Gamma(r;\beta)=
\frac{\langle \psi_\Gamma|H_g(r)|\psi_\Gamma\rangle}
{\langle \psi_\Gamma|\psi_\Gamma\rangle},
\qquad
V_\Gamma(r)=\min_\beta V_\Gamma(r;\beta).
\label{Vvar}
\end{equation}

\subsection{Kinetic energy}
Since the constituent gluon is heavy in the ILM, we can use  the nonrelativistic
expansion of the kinetic term,
\begin{equation}
\sqrt{\vec p^{\,2}+m_g^2}
=
m_g + \frac{\vec p^{\,2}}{2m_g} - \frac{\vec p^{\,4}}{8m_g^3}+\dots .
\label{kinexp}
\end{equation}
which translates to the leading average
\begin{equation}
\left\langle \sqrt{\vec p^{\,2}+m_g^2}\right\rangle
\simeq
m_g + \frac{\langle \vec p^{\,2}\rangle}{2m_g}.
\end{equation}
For the $P$-wave Gaussian we have
\begin{equation}
\langle \vec p^{\,2}\rangle_{1m} = \frac{5}{2}\beta^2,
\qquad
\langle \vec x^{\,2}\rangle_{1m} = \frac{5}{2\beta^2},
\label{momentsP}
\end{equation}
whereas for the $S$-wave Gaussian
\begin{equation}
\langle \vec p^{\,2}\rangle_{00} = \frac{3}{2}\beta^2,
\qquad
\langle \vec x^{\,2}\rangle_{00} = \frac{3}{2\beta^2}.
\label{momentsS}
\end{equation}

\subsection{Coulomb}
The Coulomb matrix element can be written as
\begin{equation}
I_C^{(\Gamma)}(r,\beta)
=
\left\langle
\frac{1}{|\vec x-\vec r/2|}
+
\frac{1}{|\vec x+\vec r/2|}
\right\rangle_\Gamma .
\label{ICdef}
\end{equation}
For the $S$-wave Gaussian we have
\begin{equation}
I_C^{(S)}(r,\beta)
=
\frac{4\beta}{\sqrt{\pi}}\,
\frac{{\rm erf}(\beta r/2)}{\beta r/2},
\label{ICS}
\end{equation}
while for the $P$-wave ansatz we obtain
\begin{align}
&I_C^{(P)}(r,\beta)
=\nonumber\\
&\frac{4\beta}{3\sqrt{\pi}}
\left[
\left(3+\frac{\beta^2 r^2}{2}\right)
\frac{{\rm erf}(\beta r/2)}{\beta r/2}
+
\frac{2}{\sqrt{\pi}}e^{-\beta^2r^2/4}
\right].
\label{ICP}
\end{align}
At small $r$, both expressions approach a constant proportional to $\beta$,
showing explicitly that the Coulomb singularity is softened by the finite size
of the gluon wavefunction. At large $r$, $I_C^{(\Gamma)}\sim 4/r$, so the
gluonic cloud resolves the two heavy sources separately.

\subsection{Confining part}
The confining expectation value is
\begin{eqnarray}
I_\sigma^{(\Gamma)}(r,\beta)
=
\left\langle
\left|\vec x-\frac{\vec r}{2}\right|
+
\left|\vec x+\frac{\vec r}{2}\right|
\right\rangle_\Gamma .
\label{Isdef}
\end{eqnarray}

{\bf S-wave result:}
The S-wave average of the confining contribution is
\begin{align}
&I_\sigma^{(S)}(r,\beta)
=\nonumber\\
&r\,\mathrm{erf}\!\left(\frac{\beta r}{2}\right)
+\frac{2}{\beta\sqrt{\pi}}\,e^{-\beta^2 r^2/4}
+\frac{2}{\beta^2 r}\,
\mathrm{erf}\!\left(\frac{\beta r}{2}\right)\,,
\label{eq:Isigma_S}
\end{align}
with the limiting behavior
\begin{align}
I_\sigma^{(S)}(r,\beta)
&=
\frac{4}{\beta\sqrt{\pi}}
+\frac{\beta r^2}{2\sqrt{\pi}}
+O(r^4), \qquad r\to 0,\nonumber\\
I_\sigma^{(S)}(r,\beta)
&=
r+\frac{2}{\beta^2 r}
+O(e^{-\beta^2 r^2/4}), \qquad r\to\infty.
\end{align}

{\bf P-wave :}
The corresponding P-wave expectation value is
\begin{align}
&I_\sigma^{(P)}(r,\beta)
=\nonumber\\
&r\,\mathrm{erf}\!\left(\frac{\beta r}{2}\right)
+\frac{2}{\beta\sqrt{\pi}},e^{-\beta^2 r^2/4}
+\frac{10}{3\beta^2 r}\,
\mathrm{erf}\!\left(\frac{\beta r}{2}\right).
\label{eq:Isigma_p}
\end{align}
with the limiting behavior 
\begin{align}
I_\sigma^{(P)}(r,\beta)
&=
\frac{16}{3\beta\sqrt{\pi}}
+\frac{2\beta r^2}{9\sqrt{\pi}}
+O(r^4), \qquad r\to 0\,\nonumber\\
I_\sigma^{(P)}(r,\beta)
&=
r+\frac{10}{3\beta^2 r}
+O(e^{-\beta^2 r^2/4}), \qquad r\to\infty.
\end{align}

Since the confining operator
\(
|\mathbf x-\mathbf r/2|+|\mathbf x+\mathbf r/2|
\)
is positive definite, the S- and P- averages are positive for small and large $r$, and their interpolations are manifestly positive for all $r>0$.

\subsection{Instanton possible term}
The instanton-induced interaction can be modeled, in the spirit of the
constituent-gluon treatment in~\cite{}, by a short-range Gaussian
centered near the midpoint,
\begin{equation}
V_{\rm inst}(\vec x)
=
- G_\Gamma\, e^{-x^2/\rho^2},
\label{Vinsthyb}
\end{equation}
where \(\rho\) is the instanton size and the coupling \(G_\Gamma\) depends on the
spin-parity channel of the light gluonic mode. Its expectation value is
\begin{equation}
\langle V_{\rm inst}\rangle_S
=
- G_S\,
\left(
\frac{\beta^2}{\beta^2+\rho^{-2}}
\right)^{3/2},
\label{VI_S}
\end{equation}
for the \(S\)-wave, and
\begin{equation}
\langle V_{\rm inst}\rangle_P
=
- G_P\,
\left(
\frac{\beta^2}{\beta^2+\rho^{-2}}
\right)^{5/2},
\label{VI_P}
\end{equation}
for the \(P\)-wave. The extra power in the \(P\)-wave case reflects the
suppression of the wavefunction at the origin. This is physically important:
instanton-induced attraction is strongest in channels with significant
short-distance overlap, and is naturally reduced for orbitally excited gluonic
configurations.

Combining these ingredients, the variational potentials are
\begin{widetext}
\begin{equation}
V_{\Sigma_g^+}(r;\beta)
=
m_g + \frac{3\beta^2}{4m_g}
-\kappa\, I_C^{(S)}(r,\beta)
+\sigma\, I_\sigma^{(S)}(r,\beta)
- G_S\left(
\frac{\beta^2}{\beta^2+\rho^{-2}}
\right)^{3/2},
\label{Vsigmag}
\end{equation}
and
\begin{equation}
V_{\Pi_u,\Sigma_u^-}(r;\beta)
=
m_g + \frac{5\beta^2}{4m_g}
-\kappa\, I_C^{(P)}(r,\beta)
+\sigma\, I_\sigma^{(P)}(r,\beta)
- G_P\left(
\frac{\beta^2}{\beta^2+\rho^{-2}}
\right)^{5/2}.
\label{Vpiu}
\end{equation}
\end{widetext}
Minimization with respect to \(\beta\) yields the adiabatic hybrid surfaces.
The splitting between \(\Pi_u\) and \(\Sigma_u^-\) is absent at \(r=0\) and arises
only through anisotropic corrections, spin-dependent interactions, and
noncentral terms beyond the spherically symmetric midpoint approximation. A
plausible parametrization of this splitting is
\begin{equation}
V_{\Sigma_u^-}(r)-V_{\Pi_u}(r)
=
\delta_2\, r^2 e^{-r^2/R_2^2},
\label{Pisplit}
\end{equation}
which vanishes quadratically as \(r\to 0\), as required by the restoration of
rotational symmetry in the gluelump limit.

One outcome of the variational analysis is the $r$-dependence of the BO potentials. Our result for the $\Sigma^-_u$ potential is compared with (the best available) lattice results in Fig.~\ref{fig_BO}, and the agreement is seen to be quite good. This suggests that BO potentials can be understood as a convolution of two Cornell potentials, corresponding to the interaction of each quark with an (octet) gluon.

A technical remark: we have minimized the Hamiltonian with respect to the parameter $\beta$ independently for each value of the quark–antiquark separation $r$. However, we find that the position of the minimum, $\beta_{\text{min}}(r)$, depends only weakly on $r$.

Furthermore, in Appendix~\ref{app_anisotropic} we performed a variational study using a Gaussian ansatz with two independent parameters, describing the transverse and longitudinal widths of the gluon distribution. Both parameters turn out to be close to the same value, $\beta_{\text{min}} \approx 0.6\,\text{GeV}$, and the resulting potentials are very similar to those obtained with the isotropic ansatz, so we do not discuss them further. This value corresponds to a root-mean-square size of the gluon distribution
$$
R_{\text{rms}}^g = \sqrt{\frac{3}{2}}\, \beta^{-1} \approx 0.40\,\text{fm}.
$$

\section{Overlaps between charmonia and hybrids} \label{sec_overlaps}
We have solved Schrodinger equation for $\Sigma_g^-$
hybrids and compare their energies to charmonium $\Sigma_g^+$, using parameterizations of potentials shown in lower Fig.\ref{fig_BO}. The first three
levels are ($(GeV))$
\be  E_{\Sigma_g^-}[n]/GeV=4.46, 4.96, 5.42,...\ee
The first two we relate to two experimentally observed states (all in $MeV$)
\ba \label{eqn_experiment}
\psi(4360), M &=& 4374\pm 7 \, , \Gamma=118\pm 12\,  \nonumber \\ 
\psi(4660), M &=& 4630\pm 6\, , \,  \Gamma=72\pm 13\,  \ea due to their decay pattern (see below) .

Note that we only use spin-independent potential, and that charmonia levels in the same approximation are 
\be  E_{\Sigma_g^+}[n]/GeV=3.10, 3.67, 4.09...
\ee
Note that the gap between $\Sigma_g^-$ and $\Sigma_g^+$ states is approximately the same for all
of them. 

\begin{figure}[h!]
    \centering
    \includegraphics[width=0.75\linewidth]{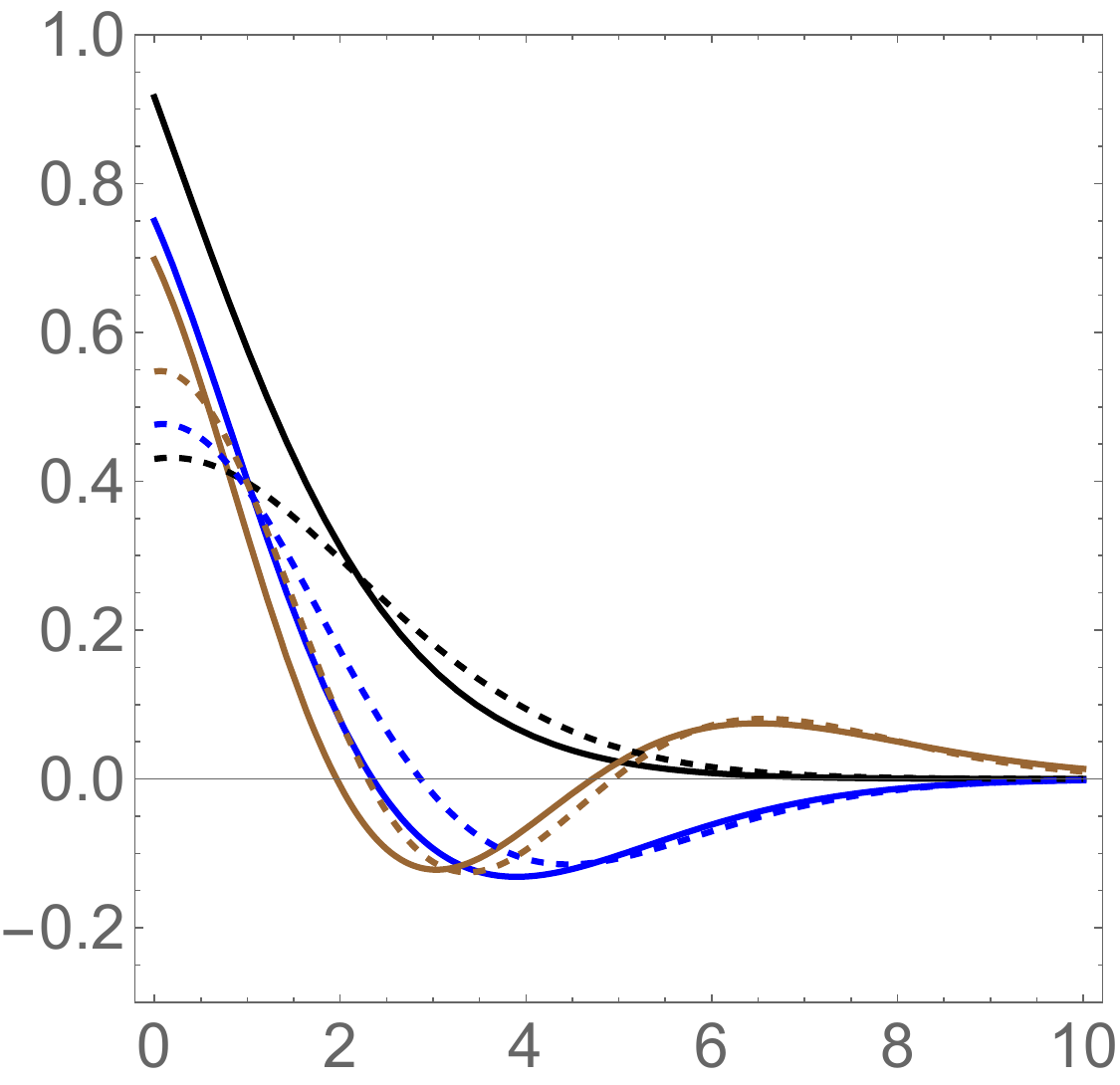}
    \caption{Three lowest normalized wave functions 
    (black,blue and brown) versus $r (GeV^{-1}) $for  $\Sigma_g^+$ charmonia (solid lines )
    compared to those for three lowest  hybrids  (dashed lines of the same colors).}
    \label{fig_WFs_Sigma_plus_minus}
\end{figure}

Let us then compare the 
normalized
wave functions
$$\phi_n(r)={\psi_n(r) \over \sqrt{\int |\psi_n)|^2 r^2 dr}} $$
for the 
$\Sigma_g^-$ potential  to those for the charmonium one $\Sigma_g^+$, see Fig.\ref{fig_WFs_Sigma_plus_minus}. 
As one can see
from it, these wave functions are very similar, except at small distances due to different Coulomb forces. 

Mentioned two hybrid candidates states were seen decaying to  1S,2S and tensor $\psi_2$
charmonium states complemented by  pair of pions, 
$\psi_{i}(3686)+\pi^+\pi^-$. Although there are no plots
to this effect known to us, we still assume that the pion pair
in $J^P=0^+$ channel is dominated by $\sigma$, the chiral
partner of pions, and treat those as two-body decays.
We further associate those with  scalar quark-gluon operator
\be O_{mix}=\sigma C (\bar c T^a \sigma_{\mu\nu} c)  G^a_{\mu\nu}\ee
where 
$C$ is unknown coupling constant and $G^a_{\mu\nu} $ is
the field strength of a gluon. Note that the bracket can be
associated with the color-magnetic moment of the $\bar c c$
pair, interacting with color-magnetic field strength of effective gluon.

Decay matrix elements should be proportional to
overlap matrix of the wave functions
\be \label{eqn_overlaps}
O_{m n}=\int  dr r^2 \phi_{\Sigma_g^-}(m)\phi_{\Sigma_g^+}(n) \ee
We calculated its $3\times 3$ upper corner with the 
obtained wave functions, the
results are given  in Table \ref{tab_overlaps}.
\begin{table}[h!]
    \centering
    \begin{tabular}{|ccc|}
    0.962506 & 0.221047 & 0.115586   \\
    0.248419 & 0.950676 & 0.126904    \\
     0.0890036 & 0.164056 & 0.972669  
    \end{tabular}
    \caption{Overlaps defined in (\ref{eqn_overlaps}).}
    \label{tab_overlaps}
\end{table}
Note that similarity of the respective wave functions leads to diagonal overlaps be all close to one, while others are much smaller. In is interesting to check if this simple idea
indeed fits the actual information about widths (\ref{eqn_experiment}). Standard expression for binary decays is
\be \Gamma_{mn}=|M_{mn}|^2 {p_{CM} \over 8\pi m_m^2}\ee
Assuming dominance of ``diagonal" decays $$\Sigma_g^-[1]\rightarrow \Sigma_g^+[1]+\sigma, \Sigma_g^-[2]\rightarrow \Sigma_g^+[2]+\sigma$$
we obtain the same $p_{CM}\approx 1\, GeV$ for both, and putting in the ratio
of masses squared one finds the state 2 more narrow than 1 (which is unusual). More precisely, using
the ratio of the observed  widths (\ref{eqn_experiment}) we get the ratio of the matrix elements
\be |{M_{22} \over M_{11} }|\approx 0.88\pm 0.09\ee
basically consistent with unity. So, our simple idea of {\em dominant and equal} diagonal $m\rightarrow m$ transitions $$ M_{mn}\sim \delta_{mn}$$ passes this  crude test. 

Of course, with more information about the decays of those states
would be needed to test it further.
Unfortunately,  branching ratios of
are not listed in RPP even when transitions themselves are ``seen". One has to wait for those to deduce empirical
information about overlap matrix elements.

\section{Spin-dependent interactions and the hybrid multiplets}

Having determined the central hybrid potentials in the constituent-gluon
molecular picture, one must next include the spin-dependent interactions. These
terms are essential because the light excitation carries nonzero angular
momentum, and the physical hybrid spectrum is organized by the coupling of the
heavy-quark spins, the orbital motion of the heavy pair, and the spin of the
constituent gluon. In the present framework the spin-dependent interactions arise
from two sources: perturbative short-distance exchange and nonperturbative
instanton-induced forces. The resulting structure is closely analogous to the
constituent-gluon Hamiltonian used for glueballs, but now with one gluon and two
heavy sources rather than two gluons.

We write the hybrid Hamiltonian in the form
\begin{equation}
H_{\rm hyb}
=
2M_Q
+
\frac{p_r^2}{M_Q}
+
V_\Gamma(r)
+
V_{\rm SD}(r),
\end{equation}
where the central adiabatic potential \(V_\Gamma(r)\) is obtained from the
variational treatment described above, and the spin-dependent contribution is
decomposed as
\begin{equation}
V_{\rm SD}(r)
=
V_{LS}^{(Q\bar Q)}(r)
+
V_{LS}^{(g)}(r)
+
V_{SS}(r)
+
V_T(r)
+
V_{\rm mix}(r).
\label{VSDdecomp}
\end{equation}
The term \(V_{LS}^{(Q\bar Q)}\) describes the coupling of the heavy-quark spin to
the orbital motion of the heavy pair, \(V_{LS}^{(g)}\) the coupling of the
constituent-gluon spin to the same orbital motion, \(V_{SS}\) the spin-spin
interaction between the heavy pair and the gluon, \(V_T\) the tensor force, and
\(V_{\rm mix}\) the noncentral interaction that mixes different BO
or partial-wave channels.

\subsection{Standard terms}

The simplest starting point is the perturbative one-gluon-exchange interaction.
Because the heavy pair is in an octet configuration inside the hybrid, the color
coefficients differ from those in ordinary quarkonium. Retaining only the
spin-dependent terms in the nonrelativistic reduction, one obtains
\begin{equation}
V_{LS}^{(Q\bar Q)}(r)
=
\frac{a_Q(r)}{M_Q^2}\,
\vec L\cdot \vec S_{Q\bar Q},
\label{VLSQQ}
\end{equation}
\begin{equation}
V_{LS}^{(g)}(r)
=
\frac{a_g(r)}{M_Q m_g}\,
\vec L\cdot \vec S_g,
\label{VLSg}
\end{equation}
\begin{equation}
V_{SS}(r)
=
\frac{b(r)}{M_Q m_g}\,
\vec S_{Q\bar Q}\cdot \vec S_g,
\label{VSSQg}
\end{equation}
and
\begin{equation}
V_T(r)
=
\frac{c(r)}{M_Q m_g}\,
S_{12}(\hat r),
\label{VTQg}
\end{equation}
where
\begin{equation}
S_{12}(\hat r)
=
3(\vec S_{Q\bar Q}\cdot \hat r)(\vec S_g\cdot \hat r)
-
\vec S_{Q\bar Q}\cdot \vec S_g .
\end{equation}

For a Coulombic interaction \(V_C(r)=-\kappa/r\), the short-distance parts of the
coefficients are
\begin{widetext}
\begin{equation}
a_Q^{(C)}(r)=\frac{\kappa}{2r^3},
\qquad
a_g^{(C)}(r)=\frac{\kappa}{r^3},
\qquad
b^{(C)}(r)=\frac{8\pi\kappa}{3}\,\delta_\Lambda^{(3)}(\vec r),
\qquad
c^{(C)}(r)=\frac{\kappa}{r^3},
\label{pertcoeff}
\end{equation}
\end{widetext}
up to channel-dependent color factors that may be absorbed into \(\kappa\). The
regulated contact term is parametrized  as
\begin{equation}
\delta_\Lambda^{(3)}(\vec r)
=
\left(\frac{\Lambda^2}{\pi}\right)^{3/2}e^{-\Lambda^2 r^2},
\label{deltareg}
\end{equation}
with \(\Lambda\sim \rho^{-1}\) of order the inverse instanton size.

The confining interaction also contributes to the spin-orbit force. For a scalar
confining potential \(V_{\rm conf}(r)\), the Thomas-precession term gives
\begin{widetext}
\begin{equation}
V_{LS}^{\rm(conf)}(r)
=
-\frac{1}{2M_Q^2r}\frac{dV_{\rm conf}}{dr}\,
\vec L\cdot \vec S_{Q\bar Q}
-\frac{1}{2M_Qm_g\,r}\frac{dV_{\rm conf}}{dr}\,
\vec L\cdot \vec S_g .
\label{VLSconf}
\end{equation}
It is convenient to combine the perturbative and confining contributions into
effective radial functions,
\begin{equation}
A_Q(r)=\frac{a_Q^{(C)}(r)}{M_Q^2}
-\frac{1}{2M_Q^2r}\frac{dV_{\rm conf}}{dr},
\qquad
A_g(r)=\frac{a_g^{(C)}(r)}{M_Qm_g}
-\frac{1}{2M_Qm_g\,r}\frac{dV_{\rm conf}}{dr},
\end{equation}
\end{widetext}
so that
\begin{equation}
V_{LS}(r)=A_Q(r)\,\vec L\cdot \vec S_{Q\bar Q}
+
A_g(r)\,\vec L\cdot \vec S_g .
\label{VLSeff}
\end{equation}

In the constituent-gluon picture there are in addition instanton-induced
short-range forces. By analogy with the glueball case, these are strongest in
channels with substantial short-distance overlap and can be represented by
Gaussian-smeared central and tensor interactions,
\begin{equation}
V_{SS}^{\rm(inst)}(r)
=
\frac{B_{\rm inst}}{M_Qm_g}\,
e^{-r^2/\rho^2}\,
\vec S_{Q\bar Q}\cdot \vec S_g,
\label{VSSinst}
\end{equation}
\begin{equation}
V_T^{\rm(inst)}(r)
=
\frac{C_{\rm inst}}{M_Qm_g}\,
\frac{1-e^{-r^2/\rho^2}}{r^3}\,
S_{12}(\hat r),
\label{VTinst}
\end{equation}
with \(\rho\) the instanton size. These terms preserve the constituent-gluon
interpretation of the light excitation while encoding the same short-distance
nonperturbative physics that generates the gluon mass and the scalar glueball
binding. The total spin-dependent interaction is then
\begin{widetext}
\begin{equation}
V_{\rm SD}(r)
=
A_Q(r)\,\vec L\cdot \vec S_{Q\bar Q}
+
A_g(r)\,\vec L\cdot \vec S_g
+
B(r)\,\vec S_{Q\bar Q}\cdot \vec S_g
+
C(r)\,S_{12}(\hat r)
+
V_{\rm mix}(r),
\label{VSDfinal}
\end{equation}
\end{widetext}
where
\begin{equation}
B(r)=\frac{8\pi\kappa}{3M_Qm_g}\,\delta_\Lambda^{(3)}(\vec r)
+\frac{B_{\rm inst}}{M_Qm_g}\,e^{-r^2/\rho^2},
\end{equation}
and
\begin{equation}
C(r)=\frac{\kappa}{M_Qm_g\,r^3}
+\frac{C_{\rm inst}}{M_Qm_g}\,\frac{1-e^{-r^2/\rho^2}}{r^3}.
\end{equation}

\subsection{Mixing terms}
The remaining term \(V_{\rm mix}(r)\) summarizes the nonadiabatic couplings that
mix nearby BO channels, most importantly the \(\Pi_u\) and
\(\Sigma_u^-\) surfaces belonging to the same \(J_g^{PC}=1^{+-}\) gluelump
multiplet. The leading symmetry-allowed form vanishes at \(r=0\) and may be
parameterized as
\begin{equation}
V_{\rm mix}(r)
=
\mu_{\Pi\Sigma}\, r\, e^{-r^2/R_{\Pi\Sigma}^2}\,{\cal O}_{\Pi\Sigma},
\label{VmixBO}
\end{equation}
where \({\cal O}_{\Pi\Sigma}\) acts in the channel space of the light-field
doublet. This term is responsible for the splitting and mixing of the
\(\Pi_u\) and \(\Sigma_u^-\) levels away from the gluelump limit.

One may now classify the resulting hybrid multiplets. For the lowest gluonic
excitation one has
\begin{equation}
J_g^{PC}=1^{+-},
\qquad
\Lambda=1,
\end{equation}
corresponding to the \(\Pi_u/\Sigma_u^-\) pair. Coupling this light excitation to
the heavy-quark spin \(S_{Q\bar Q}=0,1\) and to the orbital motion \(L\) of the
heavy pair gives the physical hybrid states. For the lowest radial level one
we use  \(L=1\), since the parity of the hybrid is
\begin{equation}
P = \epsilon\,(-1)^{L+\Lambda+1},
\qquad
C = \eta\,(-1)^{L+\Lambda+S_{Q\bar Q}},
\label{PCrule}
\end{equation}
with \(\eta=-1\) for \(u\) and \(\epsilon=\pm1\) distinguishing the reflection
quantum number in the \(\Sigma\) sector.

For \(S_{Q\bar Q}=0\), coupling \(L=1\) to \(J_g=1\) yields
\begin{equation}
J=0,1,2,
\end{equation}
and therefore the triplet
\begin{equation}
\{\,1^{-},\,0^{-+},\,1^{-+},\,2^{-+}\,\},
\end{equation}
where the \(1^{-+}\) state is spin-exotic. For \(S_{Q\bar Q}=1\), the same
construction produces the multiplet
\begin{equation}
\{\,0^{-},\,1^{-},\,2^{-},\,0^{-+},\,1^{-+},\,2^{-+},\,3^{-+}\,\}.
\end{equation}
It is convenient to reorganize these states into heavy-quark spin-symmetry
multiplets. At leading order in \(1/M_Q\), states differing only by the
orientation of \(S_{Q\bar Q}\) are degenerate, and the spin-dependent terms in
(\ref{VSDfinal}) lift this degeneracy perturbatively.

For practical calculations we will evaluate  the matrix elements of the operators
\(\vec L\cdot \vec S_{Q\bar Q}\), \(\vec L\cdot \vec S_g\),
\(\vec S_{Q\bar Q}\cdot \vec S_g\), and \(S_{12}\) in the coupled basis
\begin{equation}
\big|((L\,J_g)N\,S_{Q\bar Q})JM\big\rangle ,
\end{equation}
or equivalently in a basis where \(J_g\) and \(S_{Q\bar Q}\) are first coupled.
The diagonal spin-orbit and spin-spin matrix elements are,
\begin{equation}
\langle \vec L\cdot \vec S_{Q\bar Q}\rangle
=
\frac12\Big[J(J+1)-N(N+1)-S_{Q\bar Q}(S_{Q\bar Q}+1)\Big]
\end{equation}
after the appropriate recoupling, while the tensor operator produces off-diagonal
mixing between states with \(\Delta L=0,\pm2\). It is  this tensor force,
together with the explicit \(\Pi_u\)–\(\Sigma_u^-\) channel mixing in
(\ref{VmixBO}), that generates the fine structure of the low-lying hybrid
spectrum.


\section{$\bar c c g$ hybrids on the Light Front} \label{sec_ccg_LF}

In this section we focus on  charm–anticharm hybrids, $\bar c g c$, treated as.
 a general three-body system. It differs from, say, the $ccc$ baryon in two important ways. First, the kinematics is somewhat asymmetric, since the effective gluon mass is different,
\be 
M_g \sim 0.9\, \text{GeV} < M_c \approx 1.5\, \text{GeV}.
\ee
Second, the color structure is completely different. As discussed in \cite{Shuryak:2022thi}, baryons are approximately described by a sum of three binary potentials, whereas in the hybrid there are only two, connecting each quark to the gluon.

To recall the kinematics, we express the longitudinal momentum fractions $x_i$ in terms of Jacobi variables $\rho,\lambda$:
\ba 
x_1 &=& \frac{2 + \sqrt{6}\,\lambda + 3\sqrt{2}\,\rho}{6}, \nonumber \\
x_2 &=& \frac{2 + \sqrt{6}\,\lambda - 3\sqrt{2}\,\rho}{6}, \nonumber \\
x_3 &=& \frac{1 - \sqrt{6}\,\lambda}{3}.
\ea
These satisfy $x_1 + x_2 + x_3 = 1$.

The variables $\rho,\lambda$ are defined inside an {\em equilateral triangle} with vertices
\be 
(0, -\sqrt{2/3}), \quad (1/\sqrt{2}, 1/\sqrt{6}), \quad (-1/\sqrt{2}, 1/\sqrt{6}),
\ee
at which one of the $x_i$ equals 1.

The light-front kinetic energy is
\be \label{eqn_H_kin}
H_{\text{kin}} = \sum_i \frac{M_i^2 + p_{i,\perp}^2}{x_i},
\ee
and becomes large near the edges of the triangle, where one of the $x_i \to 0$.

As in our previous treatment of baryons on the light front \cite{Shuryak:2022thi}, we quantize the system in momentum space, where the expression above plays the role of an effective potential. The derivative terms arise from the confining interaction, which, after a standard rewriting, can be expressed as a sum of squared coordinate differences.

We use Jacobi coordinates $\vec \rho, \vec \lambda$ related to the particle coordinates $\vec r_i$, with $\sum_i \vec r_i = 0$ so that the center of mass is at rest.

For baryons, the confining potential written as a sum of relative distances reads
$$
(\vec r_1 - \vec r_2)^2 + (\vec r_1 - \vec r_3)^2 + (\vec r_3 - \vec r_2)^2
= 3(\vec\rho^{\,2} + \vec\lambda^{\,2}).
$$
In momentum representation, $\vec r_i = i\, \partial/\partial \vec p_i$, so this leads to the two-dimensional Laplacian on the triangle.

In contrast, for the hybrid (with the gluon designated as constituent number 3), we keep only two relative distances
$$
(\vec r_1 - \vec r_3)^2 + (\vec r_3 - \vec r_2)^2
= \vec\rho^{\,2} + 3\vec\lambda^{\,2},
$$
which results in a different second-order differential operator.

\begin{figure}[t]
    \centering
    \includegraphics[width=0.95\linewidth]{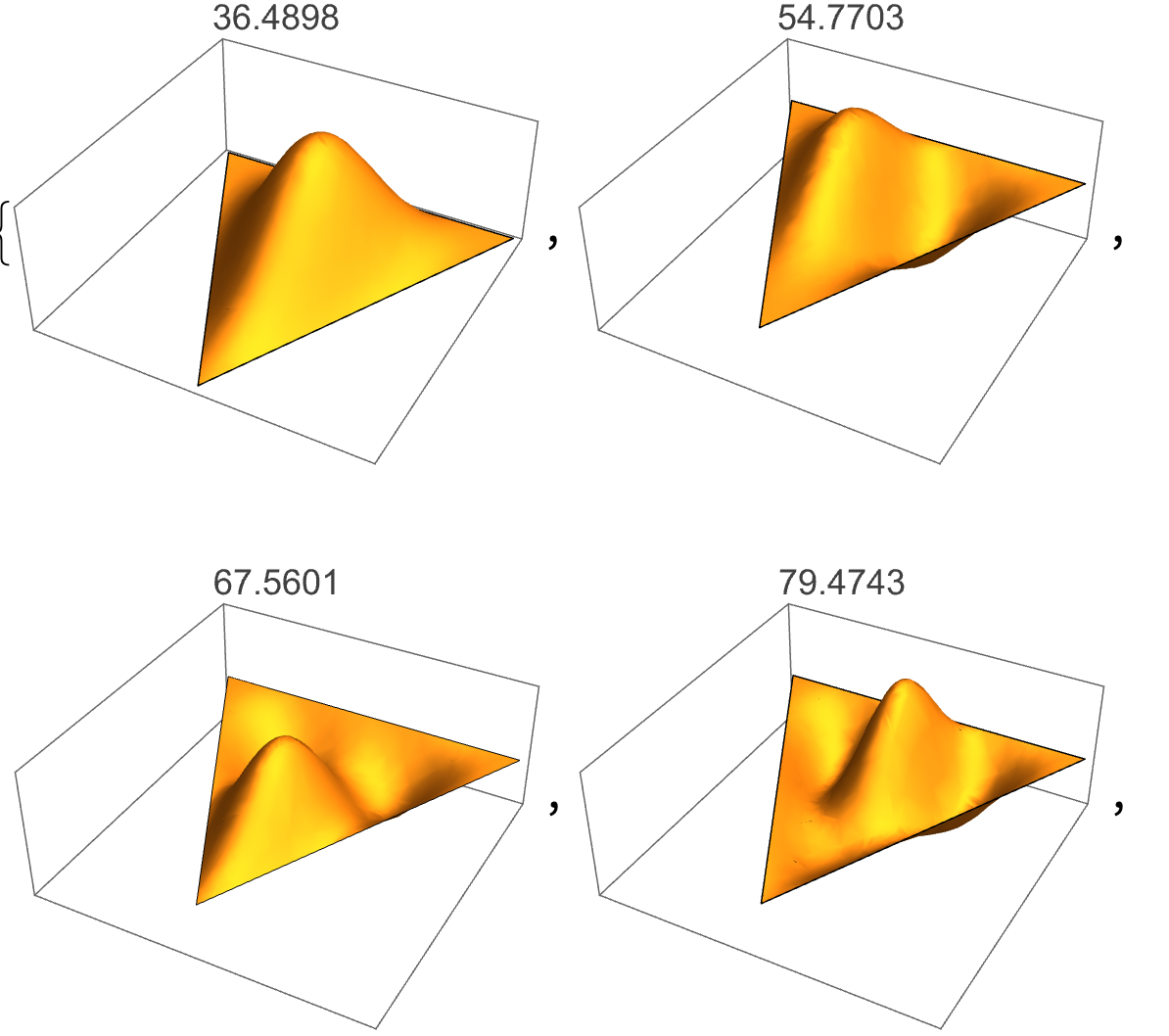}
    \caption{Four lowest eigenstates of the Hamiltonian for the $ccg$ hybrid. The numbers above the plots are eigenvalues. In contrast to the Laplacian case, there are no degenerate pairs, and the wave functions do not exhibit full triangular symmetry.}
    \label{fig_WF_ccg}
\end{figure}

We have computed the eigenfunctions of the corresponding Hamiltonian on the triangle using {\it Mathematica}. Unlike the Laplacian on this triangle (studied previously), the spectrum does not exhibit excitation brunch with double degeneracy. Four lowest eigenvalues and eigenfunctions are shown in Fig.~\ref{fig_WF_ccg}.

\begin{figure}[h!]
    \centering
    \includegraphics[width=0.85\linewidth]{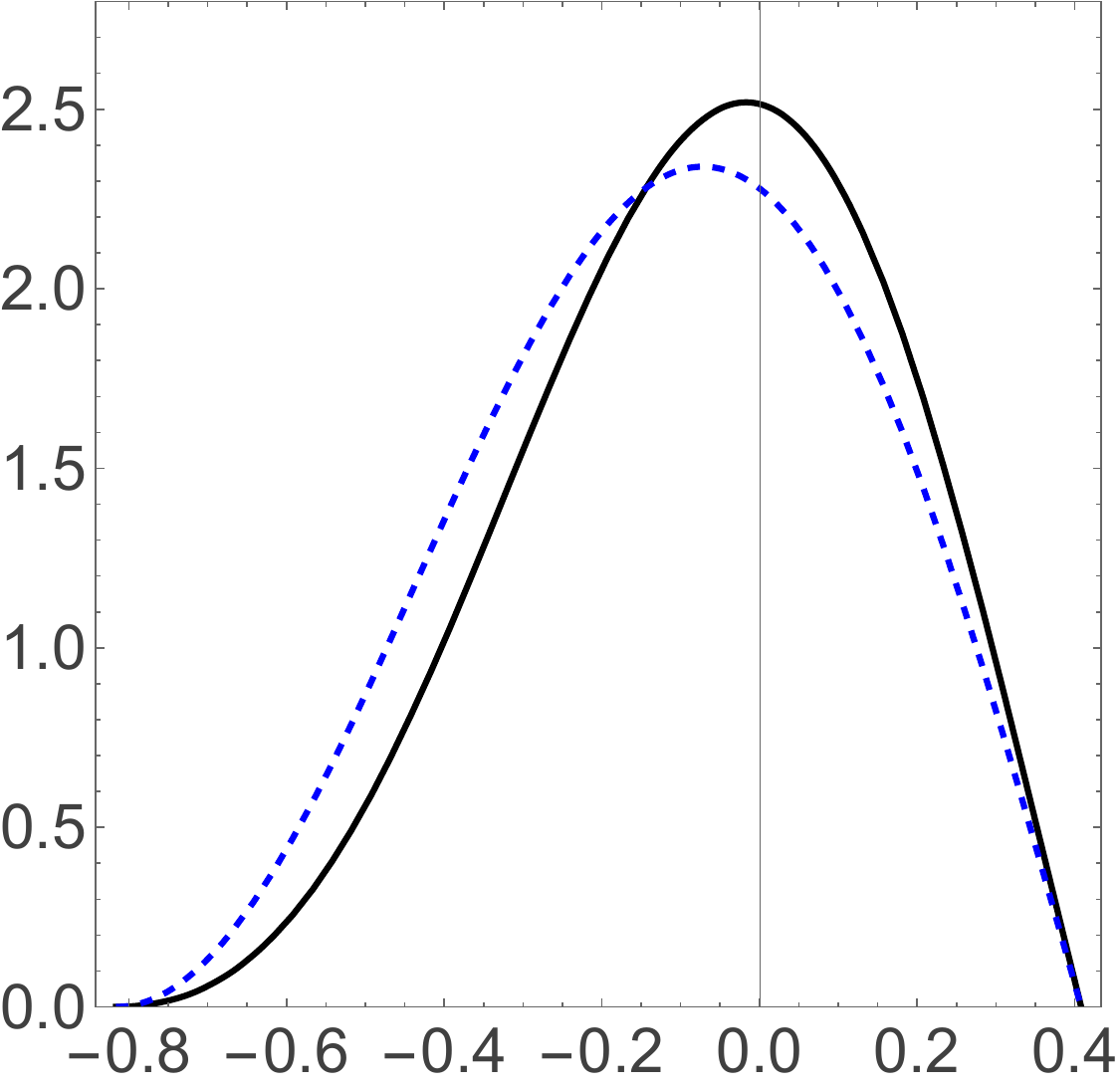}
    \caption{Ground-state wave function of the $ccg$ hybrid at $\rho=0$ as a function of $\lambda$. The solid line corresponds to the full Hamiltonian including the effective potential (\ref{eqn_H_kin}), while the dashed line shows the result without it.}
    \label{fig_WF_ccg.pdf}
\end{figure}

To illustrate the role of the ``potential'' term (\ref{eqn_H_kin}) generated by the kinetic energy, we compare the ground-state wave function obtained with (solid line) and without (dashed line) this term in Fig.~\ref{fig_WF_ccg.pdf}.

\begin{figure}[b]
    \centering
    \includegraphics[width=0.75\linewidth]{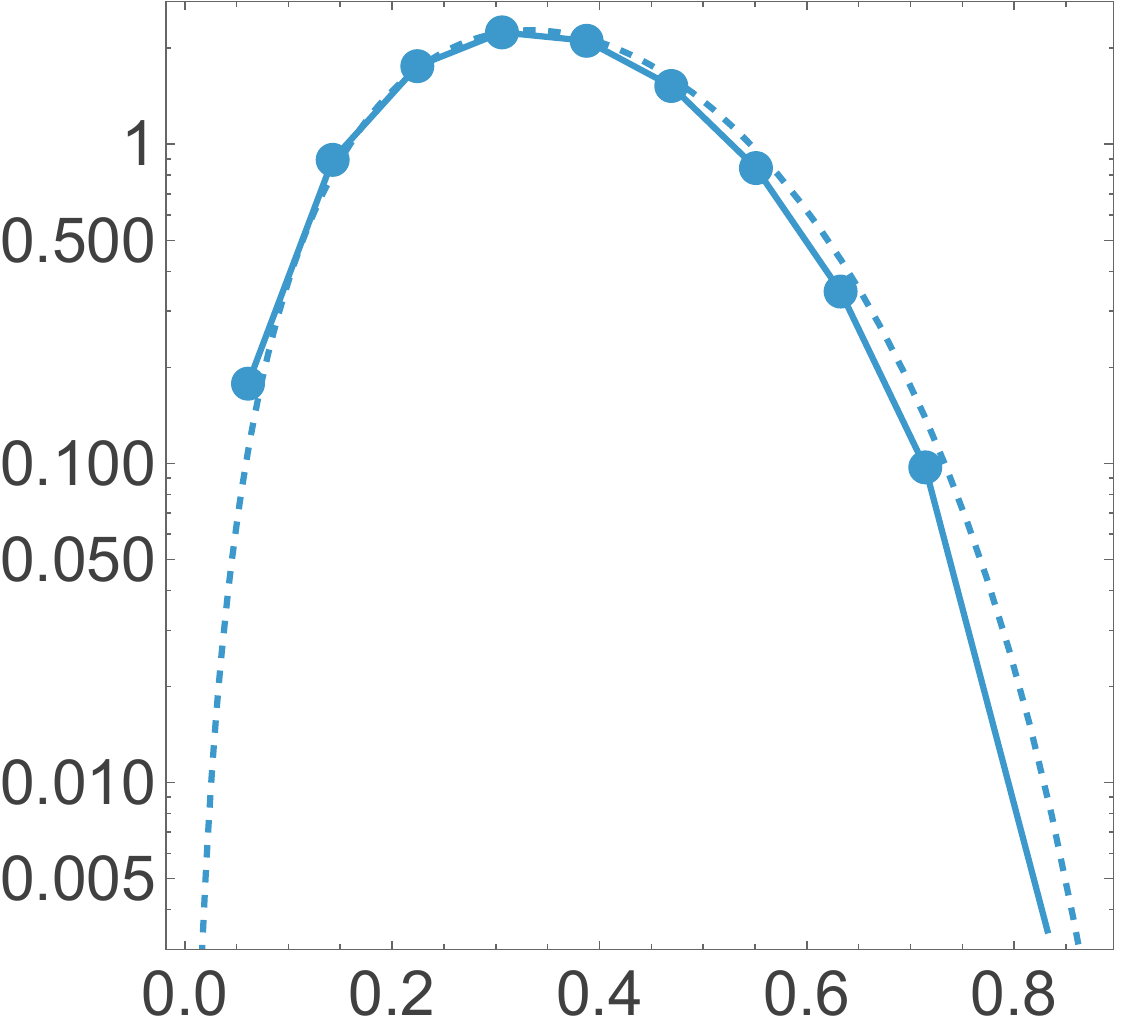}
    \caption{Gluon PDF in the $ccg$ hybrid, compared to the approximation $\sim x^3(1-x)^6$ (dashed line).}
    \label{fig_PDFg_ccg.pdf}
\end{figure}

Using the obtained ground-state wave function, we compute the gluon PDF (particle 3) by squaring it and integrating over the variable $\rho$. The result is shown in Fig.~\ref{fig_PDFg_ccg.pdf} (points and solid line).

\section{Jacobi coordinates for 4 bodies with different masses} \label{sec_qqqg}

Let us begin by commenting on the kinematic differences relative to the $ccg$ case discussed above. In addition to having one more constituent, the gluon is now not the lightest but in fact the {\em heaviest} particle, since $M_q = 350\,\text{MeV}$ while $M_g = 900\,\text{MeV}$. The gluon mass is therefore comparable to the combined mass of the three quarks. It is then natural that the gluon will mostly be at the center of the bound state. 

We start with the standard Jacobi construction for four particles with unequal masses. The coordinates are defined in terms of the original particle coordinates $x_i$ as
\ba
\alpha &=& -x_1 + x_2, \\
\beta &=& \frac{M_1 x_1 + M_2 x_2}{M_1 + M_2} - x_3, \nonumber\\
\gamma &=& \frac{M_1 x_1 + M_2 x_2 + M_3 x_3}{M_1 + M_2 + M_3} - x_4, \nonumber\\
\delta &=& \frac{M_1 x_1 + M_2 x_2 + M_3 x_3 + M_4 x_4}{M_1 + M_2 + M_3 + M_4}.
\nonumber
\ea
The coordinate $\delta$ corresponds to the center-of-mass (CM) position and is set to zero.

By computing the inverse transformation, one can express the original coordinates $x_i$ in terms of $\alpha,\beta,\gamma,\delta$, and determine the metric from the line element. The result is

\begin{widetext}
\ba
dl^2 &=&
\frac{M_1^2 + M_2^2}{(M_1 + M_2)^2}\, d\alpha^2
+ \frac{M_1^2 + 2M_1M_2 + M_2^2 + 2M_3^2}{(M_1 + M_2 + M_3)^2}\, d\beta^2
\nonumber\\
&-&
\frac{2(M_1 + M_2 - 2M_3)}{(M_1 + M_2 + M_3)(M_1 + M_2 + M_3 + M_4)}
\, d\beta \left( M_4 d\gamma + (M_1 + M_2 + M_3 + M_4)\, d\delta \right)
\nonumber\\
&+&
\frac{1}{(M_1 + M_2 + M_3 + M_4)^2}
\Big[
(M_1^2 + M_2^2 + 2M_2M_3 + M_3^2 + 2M_1(M_2 + M_3) + 3M_4^2)\, d\gamma^2
\nonumber\\
&-&
2(M_1^2 + M_2^2 + M_3^2 + 2M_2(M_3 - M_4)
+ 2M_1(M_2 + M_3 - M_4) - 2M_3M_4 - 3M_4^2)\, d\gamma\, d\delta
\nonumber\\
&+&
4(M_1 + M_2 + M_3 + M_4)^2\, d\delta^2
\Big]
\nonumber\\
&+&
\frac{2(M_1 - M_2)}{(M_1 + M_2)(M_1 + M_2 + M_3)(M_1 + M_2 + M_3 + M_4)}
\, d\alpha
\nonumber\\
&\times&
\Big[
M_3 (M_1 + M_2 + M_3 + M_4)\, d\beta
+ (M_1 + M_2 + M_3)\left(M_4 d\gamma + (M_1 + M_2 + M_3 + M_4)\, d\delta \right)
\Big].
\ea
\end{widetext}
From this metric one can construct the Laplacian in the form $g^{\mu\nu}\partial_\mu \partial_\nu$, which is non-diagonal.

We now simplify the expressions by taking equal quark masses,
\be 
M_1 = M_2 = M_3 = m, \qquad M_4 = M_g.
\ee
In this case the operator is no longer the standard Laplacian, but reduces to an anisotropic quadratic form,
\be
\frac{\alpha^2}{2} + \frac{2\beta^2}{3} + 3\gamma^2.
\ee


\section{Baryon-gluon $qqqg$ hybrid on the LF}
\label{sec:qqqg_variational_no_core}

We now discuss a  $qqqg$ hybrid state on the LF, using the same separation of variables as in the $qqq$ baryon analysis in~\cite{Shuryak:2022thi,Shuryak:2026pqt}. The transverse dynamics is described in Jacobi momenta by a harmonic-oscillator basis, while the longitudinal dynamics is described directly on the tetrahedral simplex in terms of the parton fractions $x_i$.

\subsection{Transverse wave function}
For the mass assignments
\begin{equation}
M_1=M_2=M_3\equiv m,
\qquad
M_4\equiv M_g,
\label{eq:qqqg_mass_choice_2}
\end{equation}
we use the intrinsic transverse Jacobi momenta
\begin{equation}
\bm{k}_{\alpha}=\frac{1}{\sqrt2}\left(\bm{k}_{\perp1}-\bm{k}_{\perp2}\right),
\label{eq:kalpha_2}
\end{equation}
\begin{equation}
\bm{k}_{\beta}=\sqrt{\frac23}\left(\bm{k}_{\perp3}-\frac{\bm{k}_{\perp1}+\bm{k}_{\perp2}}{2}\right),
\label{eq:kbeta_2}
\end{equation}
\begin{equation}
\bm{k}_{\gamma}=\sqrt{\frac34}\left(\bm{k}_{\perp4}-\frac{\bm{k}_{\perp1}+\bm{k}_{\perp2}+\bm{k}_{\perp3}}{3}\right),
\label{eq:kgamma_2}
\end{equation}
and identify $\bm{k}_{\gamma}$ as the gluon relative transverse coordinate to the three-quark core. 

The confining part of the baryonic hybrid is naturally taken as the sum of the three string lengths joining the gluon to the quarks. In the quadratic approximation used for the light-front oscillator problem this gives
\begin{equation}
(r_1-r_4)^2+(r_2-r_4)^2+(r_3-r_4)^2
=
\frac{\alpha^2}{2}+\frac{2\beta^2}{3}+3\gamma^2.
\label{eq:qqqg_confcoord}
\end{equation}
Therefore, in momentum representation, where $r_i\to i\,\partial/\partial p_i$, the confining contribution becomes the anisotropic second-order operator
\begin{equation}
\hat V_{\rm conf}^{\rm (LF)}
=
-\kappa\left(
\frac{1}{2}\frac{\partial^2}{\partial \alpha^2}
+\frac{2}{3}\frac{\partial^2}{\partial \beta^2}
+3\frac{\partial^2}{\partial \gamma^2}
\right),
\label{eq:qqqg_confoperator}
\end{equation}
with $\kappa$ the effective spring constant generated by the confining interaction. In contrast to the three-quark baryon, the operator is not the ordinary Laplacian: the mode $\gamma$, which describes the motion of the gluon relative to the three-quark center, is weighted more strongly than the internal quark modes $\alpha,\beta$.

To exhibit the symmetry more clearly, it is convenient to introduce canonically rescaled variables
\begin{equation}
a=-\frac{\alpha}{\sqrt{2}},\qquad
b=\sqrt{\frac{2}{3}}\,\beta,\qquad
c=\frac{\sqrt{3}}{2}\,\gamma,
\label{eq:abcvars}
\end{equation}
for which
\begin{equation}
(r_1-r_4)^2+(r_2-r_4)^2+(r_3-r_4)^2
=
a^2+b^2+4c^2,
\label{eq:qqqg_confabc}
\end{equation}
and the confining operator becomes
\begin{equation}
\hat V_{\rm conf}^{\rm (LF)}
=
-\kappa\left(
\frac{\partial^2}{\partial a^2}
+\frac{\partial^2}{\partial b^2}
+4\frac{\partial^2}{\partial c^2}
\right).
\label{eq:qqqg_confoperatorabc}
\end{equation}

When supplemented by the quadratic potentials after using the standard einbein, a complete basis set in the transverse coordinates is Gaussia. With this in mind and for the variational estimate, we take
\begin{equation}
\phi_0(\bm{k}_{\perp};b)
=
\left(\frac{1}{\pi b^2}\right)^{1/2}
\exp\!\left(-\frac{k_\perp^2}{2b^2}\right),
\label{eq:phi0qqqg_2}
\end{equation}
and
\begin{equation}
\phi_{1,m_L}(\bm{k}_{\perp};b)
=
\left(\frac{1}{\pi b^2}\right)^{1/2}
\frac{\mathcal K_{m_L}(\bm{k}_{\perp})}{b}
\exp\!\left(-\frac{k_\perp^2}{2b^2}\right),
\label{eq:phi1qqqg_2}
\end{equation}
with $m_L=\pm1$ and
\begin{equation}
\mathcal K_{+1}=k_x+i k_y,
\qquad
\mathcal K_{-1}=k_x-i k_y.
\label{eq:Kqqqg_2}
\end{equation}
The widths $b_q$ and $b_g$ are variational parameters for the spectator and gluon-core modes, respectively.

The lowest variational light-front wave function is then
\begin{widetext}
\begin{equation}
\Psi^{(0)}_{qqqg}
=
N^{(0)}_{qqqg}\,
\phi_0(\bm{k}_{\alpha};b_q)\,
\phi_0(\bm{k}_{\beta};b_q)\,
\phi_0(\bm{k}_{\gamma};b_g)\,
\varphi_4(x_1,x_2,x_3,x_g),
\label{eq:ground_final}
\end{equation}
with $\varphi_4$ given by (\ref{eq:dirichlet4_only}). The exponents are fixed by
(\ref{eq:alphas_solution}).
The lowest excited hybrid in this separable approximation is obtained by exciting the transverse gluon-core mode,
\begin{equation}
\Psi^{(1,m_L)}_{qqqg}
=
N^{(1)}_{qqqg}\,
\phi_0(\bm{k}_{\alpha};b_q)\,
\phi_0(\bm{k}_{\beta};b_q)\,
\phi_{1,m_L}(\bm{k}_{\gamma};b_g)\,
\varphi_4(x_1,x_2,x_3,x_g),
\qquad m_L=\pm 1.
\label{eq:first_excited_final}
\end{equation}
\end{widetext}
Thus the first excitation is a transverse $P$-wave in the gluon-core Jacobi coordinate, while the longitudinal structure remains the same as in the ground state.

\subsection{Longitudinal part of the LF Hamiltonian}
The light-front momentum fractions satisfy
\begin{equation}
x_i\ge 0,
\qquad
\sum_{i=1}^4 x_i=1,
\qquad
x_g\equiv x_4.
\label{eq:xi_constraint_2}
\end{equation}
In the $qqqg$ sector the longitudinal part of the LF Hamiltonian should be written in the same form as the $qqq$ baryon Hamiltonian of Ref.~\cite{Shuryak:2022thi}, but now with the correct hybrid confining topology, namely three $qg$ strings and a common einbein $a$,
\begin{widetext}
\begin{equation}
\hat H_{\parallel}(a)
=
\sum_{i=1}^{3}\frac{\omega_q}{x_i}
+
\frac{\omega_g}{x_g}
+
\sigma_{qg}\left(
3a+\frac{1}{a}\sum_{i=1}^{3}\left|\,i\frac{\partial}{\partial x_i}-i\frac{\partial}{\partial x_g}\right|^2
\right),
\label{eq:Hpar_qqqg}
\end{equation}
\end{widetext}
with
\begin{equation}
\omega_i \equiv \left\langle M_i^2+p_{i\perp}^2\right\rangle ,
\qquad
\omega_1=\omega_2=\omega_3\equiv \omega_q,
\qquad
\omega_4\equiv \omega_g .
\label{eq:omegai_def}
\end{equation}
The first term is the usual singular LF kinetic energy, while the second term is the longitudinal kinetic operator generated by the string Hamiltonian after the einbein trick. The constant term is now $3a$, reflecting the three $qg$ confining links, and the derivative operator involves only the relative quark-gluon motion. The effective string tension is correspondingly taken in the $qg$ color channel,
\begin{equation}
\sigma_{qg}=\frac{C_{qg}}{C_F}\sigma_T=\frac{9}{8}\sigma_T .
\label{eq:sigmaqg_def}
\end{equation}

We use the  trial function satisfying Dirichlet boundary
conditions at all faces of tetrahedron
\begin{equation}
\varphi_4(x_1,x_2,x_3,x_g)
=
N^{(\parallel)}_4\,
(x_1x_2x_3)^{\alpha_q}\,x_g^{\alpha_g},
\label{eq:dirichlet4_only}
\end{equation}
with 
\[
\alpha_q>\frac12,
\qquad
\alpha_g>\frac12,
\]
with the normalization unchanged,
\begin{equation}
\bigl|N_4^{(\parallel)}\bigr|^2
=
\frac{\Gamma(6\alpha_q+2\alpha_g+4)}
{\Gamma(2\alpha_q+1)^3\,\Gamma(2\alpha_g+1)}.
\label{eq:N4_dirichlet_2}
\end{equation}

For the singular LF kinetic term we still have
\begin{equation}
\left\langle \sum_{i=1}^{4}\frac{\omega_i}{x_i}\right\rangle
=
\left(6\alpha_q+2\alpha_g+3\right)
\left(
\frac{3\omega_q}{2\alpha_q}
+\frac{\omega_g}{2\alpha_g}
\right).
\label{eq:Uavg_final_revised}
\end{equation}
The longitudinal kinetic piece contributes
\begin{widetext}
\begin{equation}
\left\langle \sum_{i=1}^{3}\left|\,i\frac{\partial}{\partial x_i}-i\frac{\partial}{\partial x_g}\right|^2 \right\rangle
=
\int [dx]_4\,
\sum_{i=1}^{3}
\left|
\frac{\partial \varphi_4}{\partial x_i}
-
\frac{\partial \varphi_4}{\partial x_g}
\right|^2
\equiv {\cal K}_x(\alpha_q,\alpha_g),
\label{eq:Kx_def}
\end{equation}
and for the Dirichlet ansatz this evaluates to
\begin{equation}
{\cal K}_x(\alpha_q,\alpha_g)
=
\frac{3(\alpha_g+\alpha_q-1)(\alpha_g+3\alpha_q+1)(2\alpha_g+6\alpha_q+3)}
{(2\alpha_g-1)(2\alpha_q-1)}.
\label{eq:Kx_closed}
\end{equation}
Hence the longitudinal variational functional is
\begin{equation}
E_{\parallel}(a;\alpha_q,\alpha_g)
=
\left(6\alpha_q+2\alpha_g+3\right)
\left(
\frac{3\omega_q}{2\alpha_q}
+\frac{\omega_g}{2\alpha_g}
\right)
+
\sigma_{qg}\left[
3a+\frac{{\cal K}_x(\alpha_q,\alpha_g)}{a}
\right].
\label{eq:Epar_variational}
\end{equation}
Minimization with respect to the common einbein gives
\begin{equation}
a_\star=\sqrt{\frac{{\cal K}_x(\alpha_q,\alpha_g)}{3}},
\label{eq:a_star}
\end{equation}
and therefore
\begin{equation}
E_{\parallel}^{\rm min}(\alpha_q,\alpha_g)
=
\left(6\alpha_q+2\alpha_g+3\right)
\left(
\frac{3\omega_q}{2\alpha_q}
+\frac{\omega_g}{2\alpha_g}
\right)
+
2\sigma_{qg}\sqrt{3\,{\cal K}_x(\alpha_q,\alpha_g)}.
\label{eq:Epar_min}
\end{equation}
\end{widetext}
Unlike the previous truncated treatment, the exponents $\alpha_q$ and $\alpha_g$ are now determined by minimizing (\ref{eq:Epar_min}) numerically.

\begin{figure}
    \centering
    \includegraphics[width=0.85\linewidth]{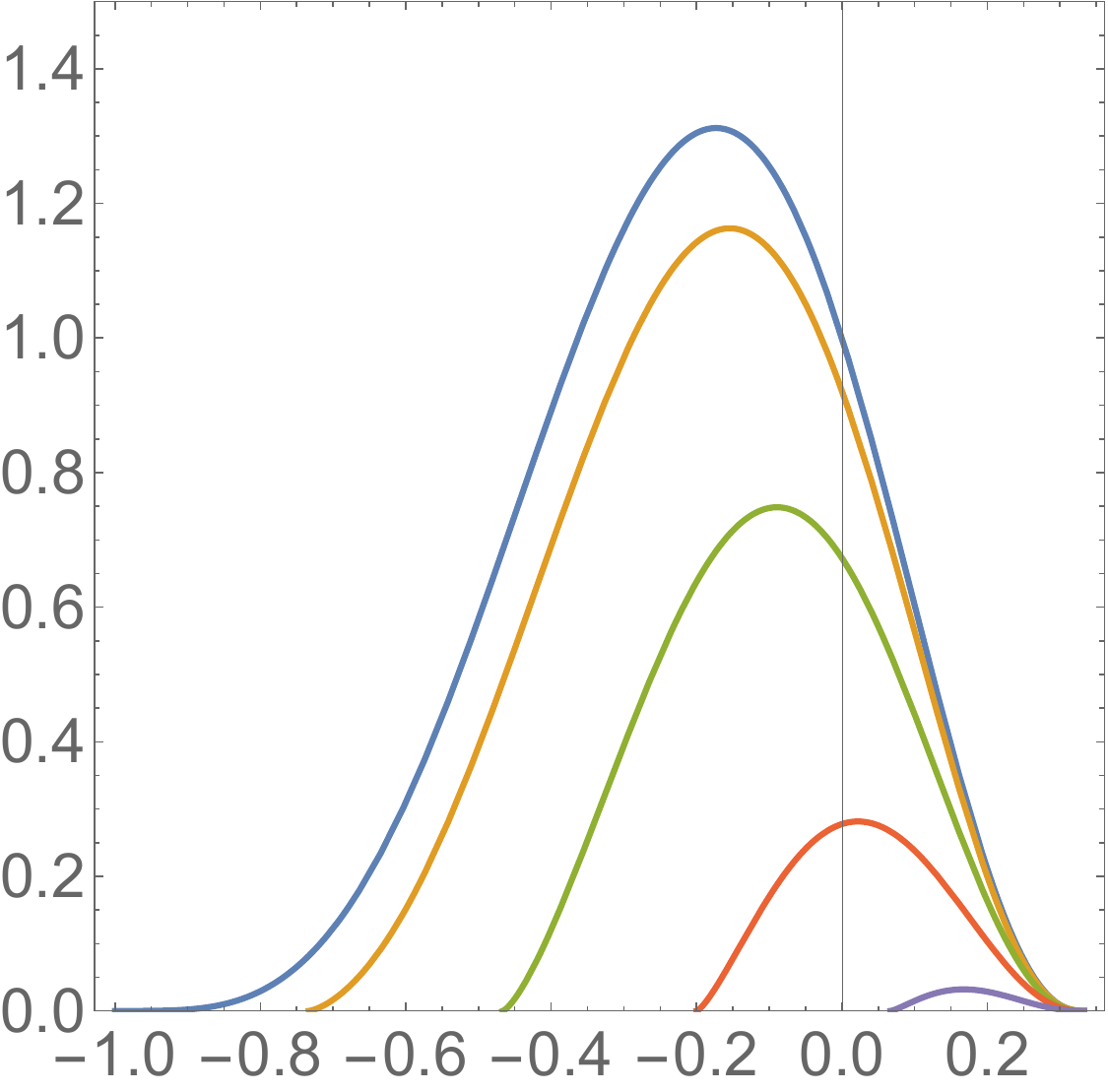}
    \caption{Variational wave function as a function of $\gamma$, at $\alpha=0$.
    Subseqquent curves, top to bottom, at $\beta=0,0.1,0.2,0.3,0.4,0.5$.}
    \label{fig_psi}
\end{figure}

Using the constituent estimate
\begin{equation}
\omega_q \simeq m^2,\qquad \omega_g \simeq M_g^2,
\end{equation}
with
\begin{equation}
m=0.35~{\rm GeV},\qquad M_g=0.90~{\rm GeV}
\end{equation}
and the string tensions
\begin{equation}
\sigma_T=0.18~{\rm GeV}^2,\qquad 
\sigma_{qg}=\frac98\,\sigma_T=0.2025~{\rm GeV}^2
\end{equation}
the numerical minimization of (\ref{eq:Epar_min}) gives
\begin{equation}
\alpha_q^{\star} = 1.365978,
\qquad
\alpha_g^{\star} = 2.505503,
\qquad
a_\star = 7.136947.
\label{eq:alpha_star_final}
\end{equation}
The shape of this wave function is illustrated in Fig.\ref{fig_psi}.
The total energy of this wave function is 
$
E_{\parallel}^{\rm min} = 13.47 $, out of which the ``cup potential" contributes $4.79$.

\subsection{Gluon PDF}
The gluon PDF is defined by integration over all but one variable
\begin{eqnarray}
&&g(x_g)
=\nonumber\\
&&\int_0^{1-x_g}dx_1\int_0^{\overline{x}_g=1-x_g-x_1}dx_2\,
\left|
\varphi_4(x_1,x_2,\overline{x}_g,x_g)
\right|^2.\nonumber\\
\label{eq:gdef_dirichlet}
\end{eqnarray}
Carrying out the simplex integral explicitly gives
\begin{equation}
g(x_g)
=
\frac{\Gamma(6\alpha_q+2\alpha_g+4)}
{\Gamma(2\alpha_g+1)\Gamma(6\alpha_q+3)}
\,
x_g^{2\alpha_g}(1-x_g)^{6\alpha_q+2}.
\label{eq:gpdf_exact}
\end{equation}
The functional form is unchanged, but the exponents are now those obtained from the minimization of the corrected longitudinal Hamiltonian. More specifically, it reads
\begin{equation}
g(x_g)\propto x_g^{5.011006}(1-x_g)^{10.195869}.
\label{eq:gpdf_numeric_corrected}
\end{equation}

\begin{figure}[h!]
    \centering
    \includegraphics[width=0.82\linewidth]{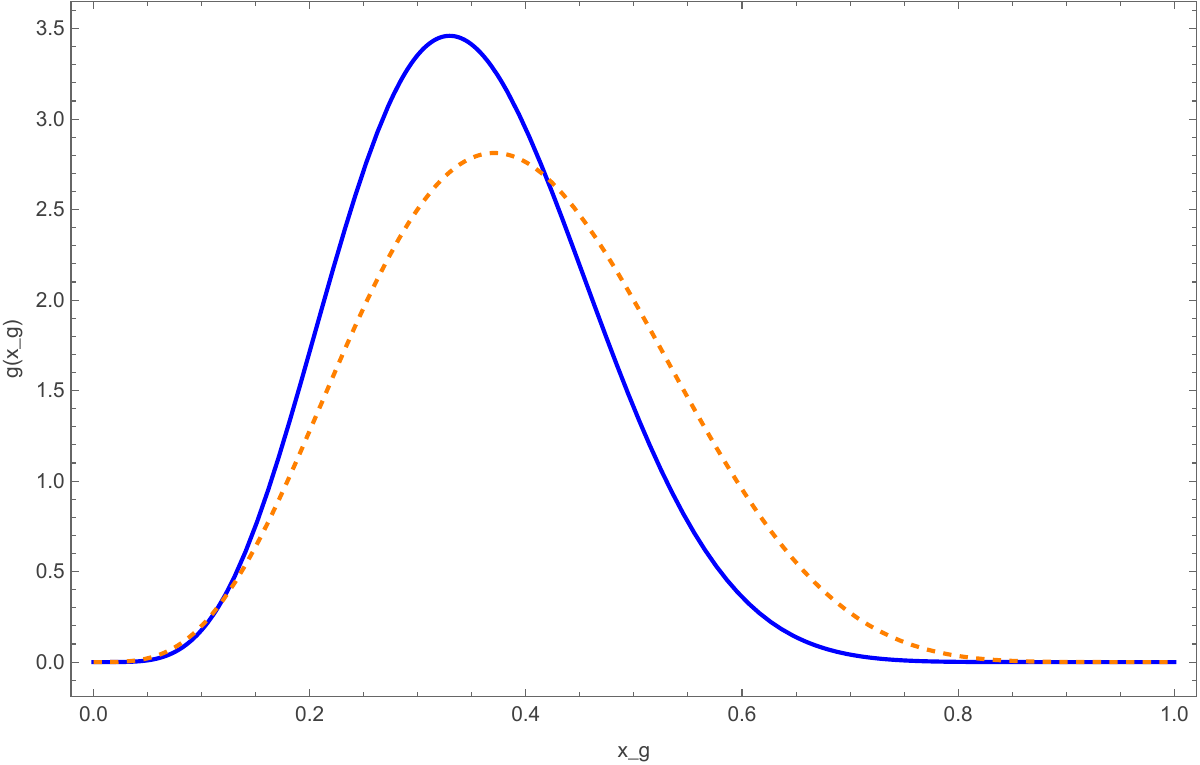}
    \caption{ Gluon PDF (solid-blue line) for the variational $qqqg$ hybrid obtained from the minimized longitudinal Hamiltonian with the LF kinetic plus confining term. The dashed-orange line is without the confining term.}
    \label{fig:gluon_pdf_qqqg_corrected}
\end{figure}

The corrected longitudinal Hamiltonian adds the derivative contribution
$2\sigma_{qg}\sqrt{3K_x(\alpha_q,\alpha_g)}$ to the variational energy. 
Expanding around the minimum of the truncated $1/x_i$ Hamiltonian gives
\begin{equation}
\delta \alpha_i
=
-\sigma_{qg}
\sum_{j=q,g}
(H^{-1})_{ij}
\left.
\frac{\partial_{\alpha_j}\!\left(3K_x\right)}{\sqrt{3K_x}}
\right|_0,
\qquad i=q,g,
\end{equation}
so the shifts of the endpoint exponents are linear in $\sigma_{qg}$ at small
string tension. The derivative term penalizes sharp longitudinal gradients in
the relative quark-gluon motion and therefore competes against the singular
$1/x_i$ piece, which by itself favors larger exponents. As a result, the
endpoint powers remain $g(x_g)\sim x_g^{2\alpha_g}$ and
$g(x_g)\sim (1-x_g)^{6\alpha_q+2}$, but the optimal values of $\alpha_q$
and $\alpha_g$ are renormalized by the longitudinal kinetic energy.

The role of the corrected longitudinal derivative term is analogous to the
longitudinal kinetic operator in BLFQ Jacobi bases: while the endpoint behavior
of the gluon PDF remains $g(x_g)\sim x_g^{2\alpha_g}(1-x_g)^{6\alpha_q+2}$,
the optimal endpoint exponents $\alpha_q$ and $\alpha_g$ are shifted because
the variational problem now balances the singular $1/x_i$ terms against a
longitudinal smoothness penalty
$\sim \sum_{i=1}^3 |\left(\partial_{x_i}-\partial_{x_g}\right)\varphi|^2$.

\subsection{Jacobi form for $qqqg$ kinematics}
Using the weighted Jacobi variables for $M_1=M_2=M_3\equiv m$ and $M_4\equiv M_g$,
\begin{widetext}
\begin{align}
\alpha &= -x_1 + x_2, \qquad
\beta = \frac{x_1+x_2}{2} - x_3, \qquad
\gamma = \frac{x_1+x_2+x_3}{3} - x_g,
\\
\delta &= \frac{m(x_1+x_2+x_3)+M_g x_g}{3m+M_g},
\end{align}
\end{widetext}
with the light-front simplex constraint
\[
x_1+x_2+x_3+x_g=1,
\]
the inverse relations for the intrinsic parton fractions are
\begin{align} \label{4body_Jacobi}
x_1 &= \frac14 - \frac{\alpha}{2} + \frac{\beta}{3} + \frac{\gamma}{4}, \\
x_2 &= \frac14 + \frac{\alpha}{2} + \frac{\beta}{3} + \frac{\gamma}{4}, \\
x_3 &= \frac14 - \frac{2\beta}{3} + \frac{\gamma}{4}, \\
x_g &= \frac14 - \frac{3\gamma}{4}.
\end{align}
We can recast the normalized longitudinal variational wave function as
\begin{widetext}
\begin{align}
\phi_{\rm var}(\alpha,\beta,\gamma)
&=
\mathcal N_4^{(\parallel)}
\Bigl(x_1 x_2 x_3\Bigr)^{\alpha_q^\star}
x_g^{\alpha_g^\star}\nonumber
\\
&=
\mathcal N_4^{(\parallel)}
\left(
\frac14 - \frac{\alpha}{2} + \frac{\beta}{3} + \frac{\gamma}{4}
\right)^{\alpha_q^\star}
\left(
\frac14 + \frac{\alpha}{2} + \frac{\beta}{3} + \frac{\gamma}{4}
\right)^{\alpha_q^\star}
\left(
\frac14 - \frac{2\beta}{3} + \frac{\gamma}{4}
\right)^{\alpha_q^\star}
\left(
\frac14 - \frac{3\gamma}{4}
\right)^{\alpha_g^\star},
\end{align}
\end{widetext}
with the variational exponents
\[
\alpha_q^\star = 1.365978, \qquad \alpha_g^\star = 2.505503,
\]
and
\[
\left|\mathcal N_4^{(\parallel)}\right|^2
=
\frac{\Gamma(6\alpha_q^\star+2\alpha_g^\star+4)}
{\Gamma(2\alpha_q^\star+1)^3\,\Gamma(2\alpha_g^\star+1)}.
\]
\begin{widetext}
The revised gluon PDF from the variational construction is

\begin{align}
g(x_g)
=
\frac{\Gamma(6\alpha_q^\star+2\alpha_g^\star+4)}
{\Gamma(2\alpha_g^\star+1)\,\Gamma(6\alpha_q^\star+3)}
\,x_g^{\,2\alpha_g^\star}(1-x_g)^{\,6\alpha_q^\star+2}
\propto x_g^{\,5.011006}(1-x_g)^{\,10.195869}.
\label{eq:gluon_pdf}
\end{align}
In terms of the Jacobi variable $\gamma$, with $x_g=(1-3\gamma)/4$, this reduces to
\begin{align}
g_\gamma(\gamma)
\equiv g\!\left(\frac{1-3\gamma}{4}\right)\left|\frac{dx_g}{d\gamma}\right|
=
\frac{3\,\Gamma(6\alpha_q^\star+2\alpha_g^\star+4)}
{4\,\Gamma(2\alpha_g^\star+1)\,\Gamma(6\alpha_q^\star+3)}
\left(\frac{1-3\gamma}{4}\right)^{2\alpha_g^\star}
\left(\frac{3(1+\gamma)}{4}\right)^{6\alpha_q^\star+2}
\propto (1-3\gamma)^{\,5.011006}(1+\gamma)^{\,10.195869}.
\end{align}

\begin{figure}[t]
    \centering
    \includegraphics[width=0.42\linewidth]{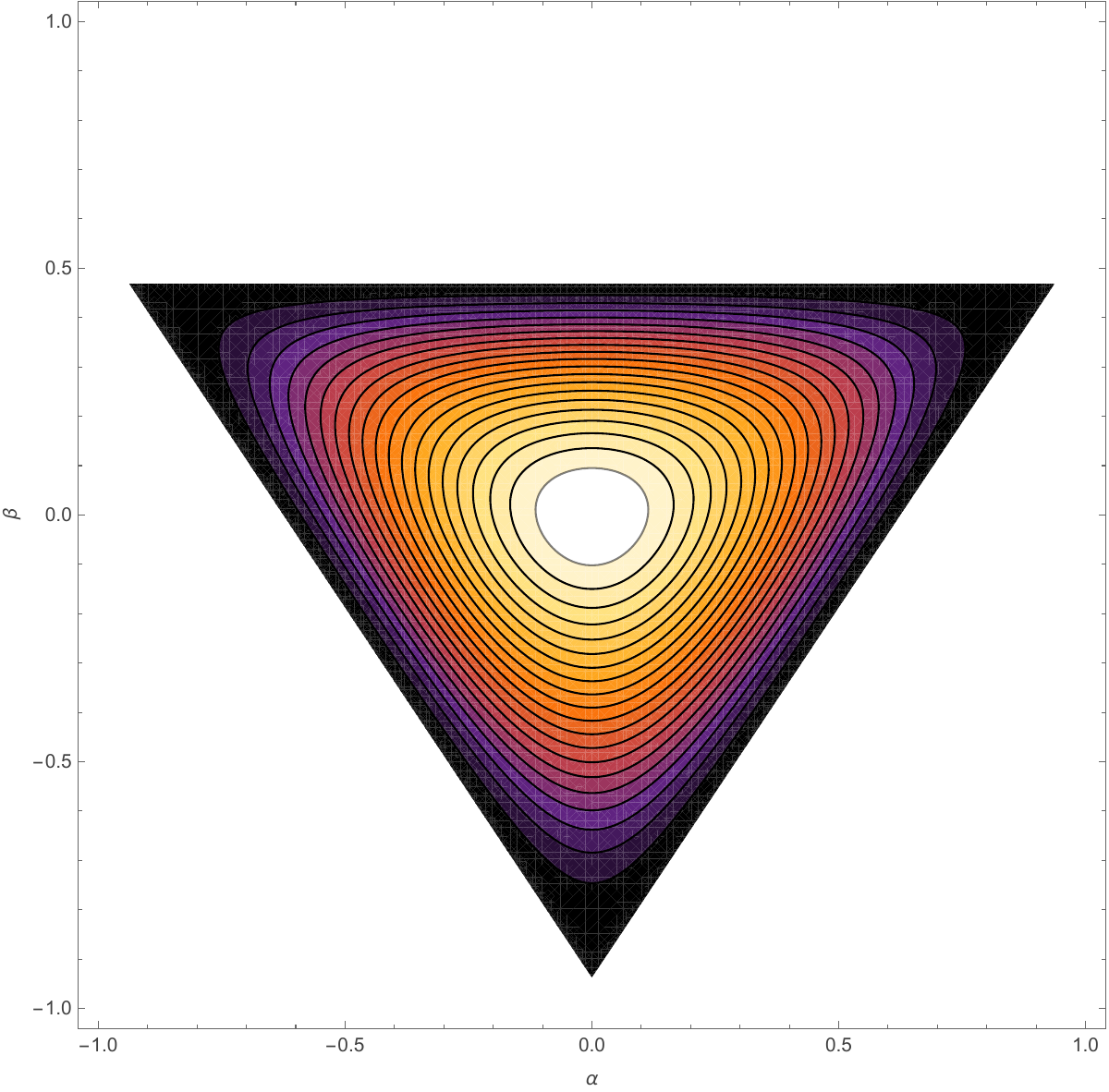}
    \caption{2D contour plot of the variational wavefunction of $qqqg$ hybrid,
    as a function of $\alpha,\beta$ variables, at $\gamma=\frac 14$.}
    \label{fig:gluon_pdf_qqqg_corrected_ab}
\end{figure}


\subsection{PDF of quarks}
For the variational longitudinal ansatz used in the $qqqg$ sector, the PDF of any one of the three identical quarks follows by integrating the squared wave function over the remaining simplex,
\begin{equation}
q(x)\equiv \int_0^{1-x}\!dx_j\int_0^{1-x-x_j}\!dx_k\,
\left|\phi_4(x,x_j,x_k,1-x-x_j-x_k)\right|^2 .
\end{equation}
Because the three quarks enter symmetrically, this gives the same distribution for $x_1$, $x_2$, or $x_3$. Using the Dirichlet integral,
\begin{equation}
q(x)=
\frac{\Gamma(6\alpha_q^\ast+2\alpha_g^\ast+4)}{\Gamma(2\alpha_q^\ast+1)\Gamma(4\alpha_q^\ast+2\alpha_g^\ast+3)}
\,x^{2\alpha_q^\ast}(1-x)^{4\alpha_q^\ast+2\alpha_g^\ast+2} .
\label{eq:qpdfx}
\end{equation}
\end{widetext}
which yields numerically
\begin{equation}
q(x)\propto x^{2.731956}(1-x)^{12.474919} .
\label{eq:qpdfxnum}
\end{equation}
This is the quark analogue of Eq.~\eqref{eq:gluon_pdf} for the gluon PDF, with the large-$x$ falloff now governed by the two spectator quarks plus the gluon.

We note that while the standard Jacobi
variable $\gamma$ isolates the gluon momentum fraction,
\begin{equation}
 x_g=\frac{1-3\gamma}{4},
\end{equation}
the single quark fraction is not aligned with one of the original $(\alpha,\beta,\gamma)$ Jacobi coordinates. For a uni-dimensional quark distribution, the pertinent choice is therefore a quark-adapted Jacobi variable obtained by singling out one quark instead of the gluon,
\begin{equation}
\gamma_q\equiv \frac{1-4x_q}{3},
\qquad
x_q=\frac{1-3\gamma_q}{4},
\label{eq:gammaqdef}
\end{equation}
with $x_q$ any one of $x_1,x_2,x_3$. In terms of $\gamma_q$, Eq.~\eqref{eq:qpdfx} becomes
\begin{widetext}
\begin{equation}
q_{\gamma_q}(\gamma_q)
\equiv q\!\left(\frac{1-3\gamma_q}{4}\right)
\bigg|\frac{d{x_q}}{{d\gamma_q}}\bigg|
=
\frac{3\,\Gamma(6\alpha_q^\ast+2\alpha_g^\ast+4)}{4\,\Gamma(2\alpha_q^\ast+1)\Gamma(4\alpha_q^\ast+2\alpha_g^\ast+3)}
\left(\frac{1-3\gamma_q}{4}\right)^{2\alpha_q^\ast}
\left(\frac{3(1+\gamma_q)}{4}\right)^{4\alpha_q^\ast+2\alpha_g^\ast+2} ,
\label{eq:qpdfgammaq}
\end{equation}
\end{widetext}
so that
\begin{equation}
q_{\gamma_q}(\gamma_q)\propto (1-3\gamma_q)^{2.731956}(1+\gamma_q)^{12.474919}.
\label{eq:qpdfgammaqnum}
\end{equation}
Equations~\eqref{eq:gammaqdef}-\eqref{eq:qpdfgammaqnum} identify the natural Jacobi variable for the quark.

\section{Light-front 
basis for $qqqg$ states with fixed $S_z$, $L_z$, and $J_z$}

In the CM frame one  classifies states by $total$ orbital and spin
quantum numbers $L,S,J$. On the light front (LF), however, 
there is no spherical symmetry, and it is natural to
construct Fock states with axial symmetry only.
This require  a basis with fixed longitudinal momentum fractions,
transverse Jacobi momenta, and fixed \emph{projections} of angular quantities
\begin{equation}
S_z,\qquad L_z,\qquad J_z=L_z+S_z.
\end{equation}
This is the basis best adapted to the LF Hamiltonian and to the gluonic transition
operator which creates one transverse gluon and excites the gluon-core relative
coordinate. The resulting states may then be recoupled, into states of good canonical $J$.

For the $qqqg$ hybrid only the three quarks are identical fermions, so the full
wave function must be antisymmetric under the permutation group $S_3$ acting on
quarks $1,2,3$. The gluon is a distinct constituent. Therefore the quark-core
orbital-spin-flavor-color part must satisfy
\begin{equation}
[\,\Psi_O^{(qqq)}\otimes \Psi_S^{(qqq)}\otimes \Psi_F^{(qqq)}\otimes \Psi_C^{(qqq)}\,]_{S_3}
=
[111].
\end{equation}
The relevant irreducible Young patterns are
\begin{equation}
[3]\;\; \ydiagram{3}, \qquad 
[21]\;\; \ydiagram{2,1}, \qquad 
[111]\;\; \ydiagram{1,1,1}.
\end{equation}

The color-singlet hybrid is obtained by coupling a color-octet $qqq$ core to the
octet gluon,
\begin{equation}
(qqq)_8\otimes g_8 \to 1_C,
\end{equation}
so the three-quark core must carry mixed permutation symmetry \[[21]_C\sim \ydiagram{2,1}\ .\]

\subsection{LF kinematics and Jacobi variables}

We use the same LF kinematics and Jacobi momenta as in the text. The parton fractions
satisfy
\begin{equation}
x_i\ge 0,\qquad x_1+x_2+x_3+x_g=1,
\end{equation}
and the intrinsic transverse Jacobi variables are
\begin{align}
\bm{k}_\alpha &=
\frac{1}{\sqrt{2}}(\bm{k}_{\perp 1}-\bm{k}_{\perp 2}),
\\
\bm{k}_\beta &=
\sqrt{\frac{2}{3}}
\left(
\bm{k}_{\perp 3}-\frac{\bm{k}_{\perp 1}+\bm{k}_{\perp 2}}{2}
\right),
\\
\bm{k}_\gamma &=
\sqrt{\frac{3}{4}}
\left(
\bm{k}_{\perp g}-\frac{\bm{k}_{\perp 1}+\bm{k}_{\perp 2}+\bm{k}_{\perp 3}}{3}
\right).
\end{align}
The variable $\bm{k}_\gamma$ describes the transverse motion of the gluon relative
to the three-quark core. Since it depends only on the quark center of mass, it is
symmetric under $S_3$. This is precisely the mode that is excited in the lowest
LF hybrid. The manuscript’s transverse basis also identifies the first hybrid
excitation as a transverse $P$-wave in the gluon-core coordinate. 

The longitudinal part is described directly on the tetrahedral simplex by a symmetric
Dirichlet-type function of the momentum fractions,
\begin{equation}
\phi_4(x_1,x_2,x_3,x_g)
\end{equation}
which is symmetric under permutations of the three quarks, i.e.\ transforms as
\[[3]\sim \ydiagram{3}\ .\] Hence the lowest LF orbital wave function is symmetric under $S_3$.

\subsection{Permutation structure and mixed $\rho,\lambda$ basis}

For the three-quark core we use the standard mixed basis associated with the two
copies of the Young pattern [21]
denoted by $\rho$ and $\lambda$.
These correspond to the two standard Young tableaux
\begin{equation}
\rho:\;
\begin{ytableau}
1 & 2 \\
3
\end{ytableau},
\qquad
\lambda:\;
\begin{ytableau}
1 & 3 \\
2
\end{ytableau}.
\end{equation}
Under exchange of quarks $1\leftrightarrow 2$,
\begin{equation}
P_{12}\,\rho=-\rho,\qquad P_{12}\,\lambda=+\lambda.
\end{equation}
The tensor product rule
\begin{equation}
[21]\otimes[21]=[3]\oplus[21]\oplus[111]
\end{equation}
implies that a mixed-symmetry color octet $[21]_C$ can combine with a mixed-symmetry
orbital-spin-flavor part $[21]_{OSF}$ to produce the required antisymmetric core
\[[111]\sim\ydiagram{1,1,1}\ .\]

\subsection{Color wave functions in the LF basis}

The two octet color cores are  denoted by ${\cal C}^{\rho,a}$ and
${\cal C}^{\lambda,a}$, with $a=1,\dots,8$ the gluon adjoint index. A convenient
explicit realization is
\begin{align}
({\cal C}^{\rho,a})_{ijk}
&=
\frac{1}{\sqrt2}\,\epsilon_{ij\ell}(t^a)_{k\ell},
\\
({\cal C}^{\lambda,a})_{ijk}
&=
\frac{1}{\sqrt6}
\Big(
\epsilon_{ik\ell}(t^a)_{j\ell}
+
\epsilon_{jk\ell}(t^a)_{i\ell}
\Big),
\end{align}
so that
\begin{equation}
P_{12}{\cal C}^{\rho,a}=-{\cal C}^{\rho,a},
\qquad
P_{12}{\cal C}^{\lambda,a}=+{\cal C}^{\lambda,a}.
\end{equation}
The two total color singlet irreducible representations of $S_3$,  are
\begin{widetext}
\begin{equation}
|{\cal C}^{A}\rangle
=
\frac{1}{\sqrt{8}}
\sum_{a=1}^8
\sum_{c_1,c_2,c_3\in\{R,G,B\}}
({\cal C}^{A,a})_{c_1 c_2 c_3}\;
|c_1\,c_2\,c_3\rangle \otimes |g^a\rangle,
\qquad A=\rho, \lambda
\end{equation}
\end{widetext}
\subsection{LF orbital states with fixed $L_z$}

On the light front, orbital structure is best specified by the transverse angular
momentum projection $L_z$. For the lowest $qqqg$ hybrid, the quark-core modes
$\bm{k}_\alpha,\bm{k}_\beta$ remain in the transverse ground state, while the
gluon-core mode $\bm{k}_\gamma$ carries one unit of transverse angular momentum.
We define
\begin{equation}
\varphi_0(\bm{k};b)=\frac{1}{\sqrt{\pi}\,b}\exp\!\left(-\frac{\bm{k}^2}{2b^2}\right),
\end{equation}
and
\begin{equation}
\varphi_{1,\ell_z}(\bm{k};b)
=
\frac{1}{\sqrt{\pi}\,b}\,
\frac{K_{\ell_z}(\bm{k})}{b}\,
\exp\!\left(-\frac{\bm{k}^2}{2b^2}\right),
\qquad
\ell_z=\pm1,
\end{equation}
with
\begin{equation}
K_{+1}=k_x+i k_y,\qquad K_{-1}=k_x-i k_y.
\end{equation}
The lowest LF orbital hybrid basis state is then
\begin{widetext}
\begin{equation}
\Phi_{L_z=\ell_z}^{\rm LF}
=
\varphi_0(\bm{k}_\alpha;b_q)\,
\varphi_0(\bm{k}_\beta;b_q)\,
\varphi_{1,\ell_z}(\bm{k}_\gamma;b_g)\,
\phi_4(x_1,x_2,x_3,x_g),
\qquad \ell_z=\pm1.
\end{equation}
\end{widetext}
Because the excited coordinate is the symmetric gluon-core mode $\gamma$, this
orbital wave function transforms as $[3]_O$ under $S_3$. The parity is
\begin{equation}
P=(-1)^{L_g+1},
\end{equation}
so the $L_z=\pm1$ gluon-core $P$-wave has positive parity and can mix with the
nucleon. 

For completeness, one may also include the unexcited orbital core
\begin{equation}
\Phi_{L_z=0}^{\rm LF,S}
=
\varphi_0(\bm{k}_\alpha;b_q)\,
\varphi_0(\bm{k}_\beta;b_q)\,
\varphi_0(\bm{k}_\gamma;b_g)\,
\phi_4(x_1,x_2,x_3,x_g),
\end{equation}
but this state does not correspond to the minimal positive-parity gluonic admixture
generated by the transverse mixing operator, to be discussed later.

\subsection{Spin basis with fixed $S_z$}

The three quarks have spin $1/2$ and the gluon has helicity/spin projection
\begin{equation}
\lambda_g=\pm1
\end{equation}
for the physical transverse gluon. The three-quark core relevant for nucleon-like
hybrids has core spin $S_q=\frac12$ and mixed permutation symmetry. We use the
standard mixed spin basis
\begin{align}
\chi^\rho_{\frac12,\frac12}
&=
\frac{1}{\sqrt2}\Big(\uparrow\downarrow\uparrow-\downarrow\uparrow\uparrow\Big),
\\
\chi^\lambda_{\frac12,\frac12}
&=
-\frac{1}{\sqrt6}\Big(\uparrow\downarrow\uparrow+\downarrow\uparrow\uparrow-2\uparrow\uparrow\downarrow\Big),
\\
\chi^\rho_{\frac12,-\frac12}
&=
\frac{1}{\sqrt2}\Big(\uparrow\downarrow\downarrow-\downarrow\uparrow\downarrow\Big),
\\
\chi^\lambda_{\frac12,-\frac12}
&=
\frac{1}{\sqrt6}\Big(\uparrow\downarrow\downarrow+\downarrow\uparrow\downarrow-2\downarrow\downarrow\uparrow\Big).
\end{align}
which satisfies
\begin{equation}
P_{12}\chi^\rho=-\chi^\rho,\qquad P_{12}\chi^\lambda=+\chi^\lambda.
\end{equation}

Instead of first coupling to good total spin $S$, on the LF we may work directly in
the product basis with fixed quark-core and gluon spin projections,
\begin{equation}
|S_z\rangle \equiv \left|\frac12,m_q\right\rangle \otimes |1,\lambda_g\rangle,
\qquad
S_z=m_q+\lambda_g.
\end{equation}
Thus the allowed values are
\begin{equation}
S_z=\pm\frac32,\ \pm\frac12.
\end{equation}
An equivalent coupled basis with good intermediate $S=\frac12,\frac32$ is obtained by
Clebsch-Gordan recoupling,
\begin{equation}
|S,S_z\rangle
=
\sum_{m_q,\lambda_g}
\left\langle \frac12\,m_q;1\,\lambda_g\middle| S\,S_z\right\rangle
\left|\frac12,m_q\right\rangle_q|1,\lambda_g\rangle_g,
\end{equation}
but for LF Hamiltonian work the uncoupled fixed-$S_z$ basis is usually more direct.

\subsection{Flavor wave functions}

For the nucleon sector with isospin $I=\frac12$ we use the standard mixed flavor
states
\begin{widetext}
\begin{align}
\phi_p^\rho &= \frac{1}{\sqrt2}(udu-duu),
&
\phi_p^\lambda &= -\frac{1}{\sqrt6}(udu+duu-2uud),
\\
\phi_n^\rho &= \frac{1}{\sqrt2}(udd-dud),
&
\phi_n^\lambda &= -\frac{1}{\sqrt6}(udd+dud-2ddu),
\end{align}
\end{widetext}
again satisfying
\begin{equation}
P_{12}\phi^\rho=-\phi^\rho,\qquad P_{12}\phi^\lambda=+\phi^\lambda.
\end{equation}

\subsection{LF spin-flavor states with fixed $S_z$}

Because the lowest LF orbital wave function is symmetric, and the color core is
mixed, the spin-flavor part of the $qqq$ core must also transform as $[21]$.
Exactly as in the equal-time construction, one may define the two mixed spin-flavor
combinations
\begin{align}
\Phi_{SF}^{\rho}(N,m_q)
&=
\frac{1}{\sqrt2}
\Big(
\phi_N^\rho \chi^\lambda_{\frac12,m_q}
+
\phi_N^\lambda \chi^\rho_{\frac12,m_q}
\Big),
\\
\Phi_{SF}^{\lambda}(N,m_q)
&=
\frac{1}{\sqrt2}
\Big(
\phi_N^\rho \chi^\rho_{\frac12,m_q}
-
\phi_N^\lambda \chi^\lambda_{\frac12,m_q}
\Big),
\end{align}
with $N=p,n$ and $m_q=\pm\frac12$.
These are the natural fixed-$m_q$ spin-flavor basis states for the quark core.

\subsection{Antisymmetric $qqq$ core on the LF}

The fully antisymmetric quark core is obtained by taking the $[111]$ component of
$[21]_C\otimes [21]_{SF}$
\begin{equation}
\Psi_{qqq}^{\rm core}(N,m_q)
=
\frac{1}{\sqrt2}
\left[
{\cal C}^{\rho}\,\Phi_{SF}^{\lambda}(N,m_q)
-
{\cal C}^{\lambda}\,\Phi_{SF}^{\rho}(N,m_q)
\right].
\label{eq:LFcore}
\end{equation}
This is the unique lowest quark-core structure compatible with the Pauli principle,
with a symmetric orbital core and with the required octet color coupling to the
gluon.

\subsection{LF OSFC basis with fixed $(L_z,S_z)$}

We can now define the fundamental light-front OSFC basis states with fixed orbital
and spin projections
\begin{widetext}
\begin{equation}
|N;L_z,S_z\rangle_{qqqg}^{\rm LF}
=
\Phi_{L_z}^{\rm LF}\,
\Psi_{qqq}^{\rm core}(N,m_q)\,
|1,\lambda_g\rangle,
\qquad
S_z=m_q+\lambda_g.
\end{equation}
or more  explicitly,
\begin{equation}
|N;L_z,S_z\rangle_{qqqg}^{\rm LF}
=
\frac{1}{\sqrt2}
\Phi_{L_z}^{\rm LF}
\left[
{\cal C}^{\rho}\,\Phi_{SF}^{\lambda}(N,m_q)
-
{\cal C}^{\lambda}\,\Phi_{SF}^{\rho}(N,m_q)
\right]
|1,\lambda_g\rangle .
\end{equation}
\end{widetext}
The allowed values for the lowest transverse hybrid are
\begin{equation}
L_z=\pm1,\qquad
m_q=\pm\frac12,\qquad
\lambda_g=\pm1,
\end{equation}
hence
\begin{equation}
S_z=\pm\frac32,\ \pm\frac12.
\end{equation}

\subsection{LF basis with fixed $J_z$}

The LF kinematical generator is
\begin{equation}
J_z=L_z+S_z.
\end{equation}
Therefore the most useful basis for diagonalization in a truncated Fock space is
often the fixed-$J_z$ basis
\begin{equation}
|N;J_z\rangle_{qqqg}^{\rm LF}
=
\sum_{L_z,m_q,\lambda_g}
C_{L_z,m_q,\lambda_g}^{(J_z)}
\,
\Phi_{L_z}^{\rm LF}\,
\Psi_{qqq}^{\rm core}(N,m_q)\,
|1,\lambda_g\rangle,
\end{equation}
subject to
\begin{equation}
J_z=L_z+m_q+\lambda_g.
\end{equation}
For the lowest hybrid with $L_z=\pm1$ the possible values are
\begin{equation}
J_z=\pm\frac52,\ \pm\frac32,\ \pm\frac12.
\end{equation}
For example, one convenient minimal basis for the nucleon-like sector with
\begin{equation}
J_z=+\frac12
\end{equation}
is
\begin{align}
|1\rangle &= \Phi_{+1}^{\rm LF}\,\Psi_{qqq}^{\rm core}(N,-\tfrac12)\,|1,-1\rangle,
\\
|2\rangle &= \Phi_{-1}^{\rm LF}\,\Psi_{qqq}^{\rm core}(N,+\tfrac12)\,|1,+1\rangle,
\\
|3\rangle &= \Phi_{+1}^{\rm LF}\,\Psi_{qqq}^{\rm core}(N,+\tfrac12)\,|1,-1\rangle,
\\
|4\rangle &= \Phi_{-1}^{\rm LF}\,\Psi_{qqq}^{\rm core}(N,-\tfrac12)\,|1,+1\rangle,
\end{align}
where the first two have $S_z=-\frac12,+\frac12$, respectively, and the last two
have $S_z=+\frac12,-\frac12$. Depending on the interaction, parity, and additional
dynamical assumptions, not all of these need contribute equally.

Similarly, for fixed
\begin{equation}
J_z=+\frac32
\end{equation}
one may use
\begin{align}
|1\rangle &= \Phi_{+1}^{\rm LF}\,\Psi_{qqq}^{\rm core}(N,+\tfrac12)\,|1,0\rangle,
\\
|2\rangle &= \Phi_{+1}^{\rm LF}\,\Psi_{qqq}^{\rm core}(N,-\tfrac12)\,|1,+1\rangle,
\\
|3\rangle &= \Phi_{-1}^{\rm LF}\,\Psi_{qqq}^{\rm core}(N,+\tfrac12)\,|1,+1\rangle,
\end{align}
if a constituent-gluon basis including the $\lambda_g=0$ polarization is retained,
or only the transverse subset if one restricts to physical gluon helicities
$\lambda_g=\pm1$.


\section{Summary and discussion}\label{sec_summary}

First, we have presented a simple variational derivation of the Born-Oppenheimer (BO) potentials for $\bar Q Q g$ hybrids, following the setup sketched in Fig.~\ref{fig_BO_schematic}. The resulting potentials, such as those shown in the upper panel of Fig.~\ref{fig_BO}, are in good agreement with lattice calculations.

We then computed the light-front (LF) wave function of the $\bar c c g$ hybrid, focusing on its longitudinal structure. As in the case of baryons, it is defined on an equilateral triangle. However, unlike baryons, it does not exhibit full symmetry, due to both the different effective gluon mass and the distinct structure of the confining interaction. The results of our variational calculation are shown in Figs.~\ref{fig_WF_ccg} and \ref{fig_WF_ccg.pdf}. The corresponding gluon PDF, i.e. the distribution over the gluon momentum fraction $x_g$, is shown in Fig.~\ref{fig_PDFg_ccg.pdf}.

We also computed the forward part of the LF wave function for the $qqqg$ hybrid, defined on a tetrahedron in Jacobi coordinates for four bodies, see (\ref{4body_Jacobi}). The variational wave function is shown in Figs.~\ref{fig_psi} and \ref{fig:gluon_pdf_qqqg_corrected_ab}. It is again used to calculate the gluon PDF, shown in Fig.~\ref{fig:gluon_pdf_qqqg_corrected}.

It is important to note that LF wave functions provide a complete description of the corresponding states. While in this work we have focused on single-particle momentum distributions (PDFs), these wave functions can be used to compute a wide range of observables, including TMDs, GPDs, form factors, and others.

As an {\em outlook for further research}, we highlight the issue of {\em nucleon–$qqqg$ hybrid mixing}. This should ultimately be connected to the gluon PDF extracted from experiments. The connection is nontrivial, since experimental PDFs are measured at relatively large momentum scales, $\mu \sim 3$–$10\,\text{GeV}$. These are typically evolved downward using perturbative (DGLAP) equations to a so-called ``bridging scale'' $\mu \sim 1\,\text{GeV}$, where they are expected to match hadronic quark–gluon models.

However, those evolution equations were derived under assumptions that are not valid in this regime. In particular, they treat gluons as soft, massless, and emitted without further interactions with quarks. These assumptions break down at scales comparable to the effective gluon mass. A correct evolution framework should smoothly connect perturbative descriptions to a spectroscopic picture of $qqqg$ hybrids, such as those developed in this work. This should be part of a more general renormalization group approach, defined through a gradual increase of the Fock-space content of the wave functions. We hope to address these issues in future work.

{\centerline{\bf Acknowledgements}}
\vskip 0.5cm
This work is supported by the Office of Science, U.S. Department of Energy under Contract  No. DE-FG-88ER40388.
This research is also supported in part within the framework of the Quark-Gluon Tomography (QGT) Topical Collaboration, under contract no. DE-SC0023646.

\appendix

\section{The variational BO potential for the $\bar Q Q g$ hybrids}
At short distances the interaction is controlled by one-gluon exchange,
\begin{equation}
V_C(r)=\frac{\alpha_s}{r}\,T_1\!\cdot T_2
=
-\frac{\alpha_s}{2r}\Big(C_{R_1}+C_{R_2}-C_R\Big).
\end{equation}
Thus for ordinary quarkonium, $3\otimes \bar 3\to 1$, one recovers the
attractive singlet Coulomb term
\begin{equation}
V_C^{(1)}(r)=-\frac{4}{3}\frac{\alpha_s}{r},
\end{equation}
whereas for the octet heavy-quark pair relevant to hybrids,
$3\otimes \bar 3\to 8$, we have
\begin{equation}
V_C^{(8)}(r)=+\frac{1}{6}\frac{\alpha_s}{r},
\end{equation}
which is repulsive. By contrast, for two adjoint sources projected onto a
singlet, $8\otimes 8\to 1$, as in a constituent two-gluon glueball model,
the Coulomb interaction is
\begin{equation}
V_C^{(8\otimes 8\to 1)}(r)=-3\,\frac{\alpha_s}{r},
\end{equation}
namely $9/4$ times stronger than the quarkonium singlet coefficient.

Assuming Casimir scaling, the adjoint string tension is related to the
fundamental one by
\begin{equation}
\sigma_{\rm adj}=\frac{C_A}{C_F}\sigma_F=\frac{9}{4}\sigma_F
\qquad (SU(3)),
\end{equation}
although for adjoint sources this linear behavior is expected to be screened
at sufficiently large distances by gluon pair creation.

Although the heavy quark pair is in a color-octet configuration at short
distances, this does mean that the large-distance behavior of the
BO potential is governed by an adjoint string tension.
At small $r$, the $Q\bar Q$ pair acts as a compact octet source, leading
to the repulsive Coulomb interaction.
As the separation increases, the system reorganizes into a
color-singlet three-body configuration $Q\bar Q g$, in which the gluonic
degree of freedom screens the octet charge.
The resulting flux configuration is not an adjoint string between the
heavy quarks, but rather a string with a dynamical gluonic excitation.
As a result, the hybrid BO potentials behave asymptotically as
\begin{equation}
V_\Gamma(r) \simeq \sigma_F r + \frac{C_L}r +\cdots ,
\end{equation}
with the gluonic excitation contributing only to the subleading structure
of the potential, by modifying the coefficient $C_L$ of the Luscher term.


For clarity, we briefly recall the definition of the quadratic Casimir
operator. For a representation $R$ of $SU(N_c)$, the generators $T^a_R$
satisfy
\begin{equation}
T^a_R T^a_R = C_R\, \mathbf{1},
\end{equation}
where $C_R$ is the quadratic Casimir invariant of the representation.
It characterizes the strength of color interactions for a given color
charge.

For the representations relevant to QCD with $N_c=3$, the relevant Casimirs are
\begin{equation}
C_F = \frac{N_c^2-1}{2N_c} = \frac{4}{3}
\qquad \text{(fundamental, } 3),
\end{equation}
\begin{equation}
C_A = N_c = 3
\qquad \text{(adjoint, } 8),
\end{equation}
and
\begin{equation}
C_1 = 0
\qquad \text{(singlet)}.
\end{equation}

More generally, for two color sources in representations $R_1$ and $R_2$
combined into a channel $R$, the color interaction is governed by
\begin{equation}
T_1 \cdot T_2
=
\frac{1}{2}\left(C_R - C_{R_1} - C_{R_2}\right),
\label{CasimirIdentity}
\end{equation}
which directly determines the Coulomb potential at short distances,
\begin{equation}
V_C(r)
=
\frac{\alpha_s}{r}\, T_1 \cdot T_2.
\end{equation}

For a quark-gluon system,
\begin{equation}
3 \otimes 8 = 3 \oplus \bar{6} \oplus 15,
\end{equation}
and the energetically favored channel in the hybrid corresponds to the
triplet. Using $C_F=4/3$ and $C_A=3$, one finds
\begin{equation}
T_Q \cdot T_g
=
\frac{1}{2}\left(C_3 - C_3 - C_8\right)
=
-\frac{C_A}{2}
=
-\frac{3}{2}.
\end{equation}
It is therefore convenient to define
\begin{equation}
V_{Qg}(r) = -\frac{C_{Qg}\,\alpha_s}{r},
\qquad
C_{Qg} = \frac{C_A}{2} = \frac{3}{2}.
\end{equation}
Thus the quark-gluon interaction is attractive and stronger than the
fundamental $Q\bar Q$ singlet Coulomb interaction by a factor $9/8$.

\section{Molecular labeling}

The labels $\Sigma_u^-$ and $\Pi_u$ used for hybrid BO
potentials follow the standard spectroscopic notation of diatomic
molecules, reflecting the cylindrical symmetry of a static $Q\bar Q$
pair. The relevant symmetry group is  the group of
rotations around the interquark axis combined with discrete
symmetries.

The symbol $\Lambda=\Sigma,\Pi,\Delta,\dots$ denotes the magnitude of
the projection of the total angular momentum of the light degrees of
freedom onto the molecular axis,
\begin{equation}
\Lambda = |\hat r \cdot \vec J_g|,
\end{equation}
with $\Lambda=0,1,2,\dots$ corresponding to $\Sigma,\Pi,\Delta,\dots$
respectively.
The subscript $g$ or $u$ specifies the eigenvalue under the combined
operation of charge conjugation and inversion through the midpoint,
\begin{equation}
(CP)_{\rm light} = \pm 1,
\end{equation}
with $g=+1$ (even)  and $u=-1$ (odd).

For $\Sigma$ states ($\Lambda=0$) there is an additional quantum number
$\epsilon=\pm$ that distinguishes the behavior under reflection in a
plane containing the molecular axis. The superscript $+$ or $-$ labels
this reflection symmetry. $\Pi_u$ corresponds to a gluonic excitation with one unit of
angular momentum about the interquark axis and odd $(CP)$ quantum
number, while $\Sigma_u^-$ denotes a state with zero angular momentum
projection, odd $(CP)$, and negative reflection parity.

These quantum numbers originate from the symmetry of the static
quark-antiquark system and provide a complete classification of the
adiabatic hybrid potentials, in direct analogy with electronic states
in ordinary diatomic molecules.

\section{$I_\sigma^{(P)}(r,\beta)$, explicit averaging over gluon location}

Here we give the details for the P-wave expectation value
\begin{equation}
I_\sigma^{(P)}(r,\beta)
=
\left\langle
\left|\mathbf x-\frac{\mathbf r}{2}\right|
+
\left|\mathbf x+\frac{\mathbf r}{2}\right|
\right\rangle_{P},
\end{equation}
using the Gaussian ansatz
\begin{equation}
\psi_{1m}(\mathbf x)=N_1\,x\,e^{-\beta^2 x^2/2}Y_{1m}(\hat x).
\end{equation}
The normalization condition gives
\begin{equation}
1=N_1^2 \int d^3x,x^2 e^{-\beta^2 x^2}|Y_{1m}|^2
= N_1^2 \int_0^\infty dx,x^4 e^{-\beta^2 x^2},
\end{equation}
hence
\begin{equation}
N_1^2=\frac{8\beta^5}{3\sqrt{\pi}}.
\end{equation}

By symmetry,
\begin{equation}
I_\sigma^{(P)}(r,\beta)
=2N_1^2 \int_0^\infty dx\,x^4 e^{-\beta^2 x^2}\,A(x,r),
\end{equation}
with
\begin{equation}
A(x,r)=\frac12\int_{-1}^{1} d\mu,
\sqrt{x^2+\frac{r^2}{4}-xr\mu}.
\end{equation}
This integral evaluates to the piecewise form
\begin{equation}
A(x,r)=
\begin{cases}
\dfrac r2+\dfrac{2x^2}{3r}, & x<\frac r2,\\
x+\dfrac{r^2}{12x}, & x>\frac r2.
\end{cases}
\end{equation}
hence
\begin{align}
I_\sigma^{(P)}(r,\beta)
&=
N_1^2 \Bigg[
\int_0^{r/2} dx,x^4 e^{-\beta^2 x^2}
\left(r+\frac{4x^2}{3r}\right)
\nonumber\\
&\quad\quad\quad+
\int_{r/2}^\infty dx,x^4 e^{-\beta^2 x^2}
\left(2x+\frac{r^2}{6x}\right)
\Bigg].
\end{align}
Using standard Gaussian integrals, we obtain
\begin{equation}
I_\sigma^{(P)}(r,\beta)
=
r\,\mathrm{erf}\!\left(\frac{\beta r}{2}\right)
+\frac{2}{\beta\sqrt{\pi}}e^{-\beta^2 r^2/4}
+\frac{10}{3\beta^2 r},
\mathrm{erf}\!\left(\frac{\beta r}{2}\right).
\end{equation}
which is the result in~\eqref{eq:Isigma_p}.

\section{Anisotropic variational ansatze}
\label{app_anisotropic}

The variational construction of Sec.~III uses a spherically symmetric Gaussian ansatz centered at the midpoint between the heavy sources. This choice is analytically convenient, but at fixed interquark separation $r$ the light-sector Hamiltonian has only cylindrical symmetry around the molecular axis. The reduction from spherical to axial symmetry is especially transparent in the confining term
\begin{equation}
V_{\rm conf}(\vec x;r)=\sigma\left(\left|\vec x-\frac{\vec r}{2}\right|+\left|\vec x+\frac{\vec r}{2}\right|\right),
\end{equation}
with $\vec r=r\,\hat z$. Along the molecular axis one has $\rho=0$, and for $|z|<r/2$ the sum of distances is exactly constant,
\begin{equation}
\left|\vec x-\frac{\vec r}{2}\right|+\left|\vec x+\frac{\vec r}{2}\right|=r.
\end{equation}
Thus the confining interaction develops a flat valley between $Q$ and $\bar Q$, suggesting that the gluonic wavefunction should be allowed to spread differently along the longitudinal and transverse directions. This observation motivates a two-parameter extension of the variational ansatz.

We therefore introduce cylindrical coordinates $(\rho,\phi,z)$ with the $z$ axis chosen along the heavy-quark separation. For the $\Pi_u$ doublet, with $\Lambda=1$, the simplest anisotropic Gaussian ansatz is
\begin{equation}
\psi_{\Pi}^{(m=\pm 1)}(\rho,\phi,z)
=
N_{\Pi}\,\rho\,e^{im\phi}
\exp\!\left(-\frac{\rho^2}{2a_\perp^2}-\frac{z^2}{2a_\parallel^2}\right),
\end{equation}
with $m=\pm 1$, while for the $\Sigma_u^-$ partner one may use
\begin{equation}
\psi_{\Sigma}(\rho,\phi,z)
=
N_{\Sigma}\,z
\exp\!\left(-\frac{\rho^2}{2a_\perp^2}-\frac{z^2}{2a_\parallel^2}\right).
\end{equation}
Here $a_\perp$ and $a_\parallel$ are independent variational widths describing the transverse and longitudinal extent of the constituent-gluon wavefunction. Direct integration gives
\begin{equation}
N_{\Pi}^2=\frac{1}{\pi^{3/2}a_\perp^4 a_\parallel},
\qquad
N_{\Sigma}^2=\frac{2}{\pi^{3/2}a_\perp^2 a_\parallel^3}.
\end{equation}

\subsection{Kinetic}
The kinetic expectation values are most easily evaluated from $\langle \vec p^{\,2}\rangle=\int d^3x\, |\nabla \psi|^2$. One finds
\begin{equation}
\langle \vec p^{\,2}\rangle_{\Pi}
=
\frac{2}{a_\perp^2}+\frac{1}{2a_\parallel^2},
\qquad
\langle \vec p^{\,2}\rangle_{\Sigma}
=
\frac{1}{a_\perp^2}+\frac{3}{2a_\parallel^2},
\end{equation}
and therefore
\begin{align}
&T_{\Pi}
=
 m_g+\frac{1}{2m_g}\left(\frac{2}{a_\perp^2}+\frac{1}{2a_\parallel^2}\right),
\nonumber\\
&T_{\Sigma}
=
 m_g+\frac{1}{2m_g}\left(\frac{1}{a_\perp^2}+\frac{3}{2a_\parallel^2}\right).
\end{align}

\subsection{Coulomb}
The Coulomb matrix elements are
\begin{equation}
I_C^{(\Pi)}(r;a_\perp,a_\parallel)
=
\int d^3x\,|\psi_{\Pi}(\vec x)|^2
\left(
\frac{1}{|\vec x-\vec r/2|}+\frac{1}{|\vec x+\vec r/2|}
\right),
\end{equation}
\begin{equation}
I_C^{(\Sigma)}(r;a_\perp,a_\parallel)
=
\int d^3x\,|\psi_{\Sigma}(\vec x)|^2
\left(
\frac{1}{|\vec x-\vec r/2|}+\frac{1}{|\vec x+\vec r/2|}
\right),
\end{equation}
which, after the trivial $\phi$ integration, reduce to the two-dimensional forms
\begin{widetext}
\begin{equation}
I_C^{(\Pi)}(r;a_\perp,a_\parallel)
=
\frac{2}{\sqrt{\pi}\,a_\perp^4 a_\parallel}
\int_0^{\infty} d\rho\,\rho^3 e^{-\rho^2/a_\perp^2}
\int_{-\infty}^{\infty} dz\,e^{-z^2/a_\parallel^2}
\left(
\frac{1}{\sqrt{\rho^2+(z-r/2)^2}}+
\frac{1}{\sqrt{\rho^2+(z+r/2)^2}}
\right),
\end{equation}
\begin{equation}
I_C^{(\Sigma)}(r;a_\perp,a_\parallel)
=
\frac{4}{\sqrt{\pi}\,a_\perp^2 a_\parallel^3}
\int_0^{\infty} d\rho\,\rho e^{-\rho^2/a_\perp^2}
\int_{-\infty}^{\infty} dz\,z^2 e^{-z^2/a_\parallel^2}
\left(
\frac{1}{\sqrt{\rho^2+(z-r/2)^2}}+
\frac{1}{\sqrt{\rho^2+(z+r/2)^2}}
\right).
\end{equation}

\subsection{Confining}
The confining matrix elements are defined analogously
\begin{equation}
I_\sigma^{(\Pi)}(r;a_\perp,a_\parallel)
=
\int d^3x\,|\psi_{\Pi}(\vec x)|^2
\left(\left|\vec x-\frac{\vec r}{2}\right|+\left|\vec x+\frac{\vec r}{2}\right|\right),
\end{equation}
\begin{equation}
I_\sigma^{(\Sigma)}(r;a_\perp,a_\parallel)
=
\int d^3x\,|\psi_{\Sigma}(\vec x)|^2
\left(\left|\vec x-\frac{\vec r}{2}\right|+\left|\vec x+\frac{\vec r}{2}\right|\right).
\end{equation}
After the angular integration these become
\begin{equation}
I_\sigma^{(\Pi)}(r;a_\perp,a_\parallel)
=
\frac{2}{\sqrt{\pi}\,a_\perp^4 a_\parallel}
\int_0^{\infty} d\rho\,\rho^3 e^{-\rho^2/a_\perp^2}
\int_{-\infty}^{\infty} dz\,e^{-z^2/a_\parallel^2}
\left(
\sqrt{\rho^2+(z-r/2)^2}+
\sqrt{\rho^2+(z+r/2)^2}
\right),
\end{equation}
\begin{equation}
I_\sigma^{(\Sigma)}(r;a_\perp,a_\parallel)
=
\frac{4}{\sqrt{\pi}\,a_\perp^2 a_\parallel^3}
\int_0^{\infty} d\rho\,\rho e^{-\rho^2/a_\perp^2}
\int_{-\infty}^{\infty} dz\,z^2 e^{-z^2/a_\parallel^2}
\left(
\sqrt{\rho^2+(z-r/2)^2}+
\sqrt{\rho^2+(z+r/2)^2}
\right).
\end{equation}

\subsection{Instanton}
For a Gaussian instanton-induced interaction,
\begin{equation}
V_{\rm inst}(\vec x)=-G_\Gamma e^{-x^2/\rho_0^2},
\end{equation}
with $\rho_0$ the instanton size, the expectation values are elementary. For the $\Pi_u$ ansatz one obtains
\begin{equation}
\mathcal I_{\Pi}(a_\perp,a_\parallel)
=
\int d^3x\,|\psi_{\Pi}(\vec x)|^2 e^{-x^2/\rho_0^2}
=
\left(\frac{a_\perp^{-2}}{a_\perp^{-2}+\rho_0^{-2}}\right)^2
\left(\frac{a_\parallel^{-2}}{a_\parallel^{-2}+\rho_0^{-2}}\right)^{1/2},
\end{equation}
while for the $\Sigma_u^-$ ansatz one finds
\begin{equation}
\mathcal I_{\Sigma}(a_\perp,a_\parallel)
=
\int d^3x\,|\psi_{\Sigma}(\vec x)|^2 e^{-x^2/\rho_0^2}
=
\left(\frac{a_\perp^{-2}}{a_\perp^{-2}+\rho_0^{-2}}\right)
\left(\frac{a_\parallel^{-2}}{a_\parallel^{-2}+\rho_0^{-2}}\right)^{3/2}.
\end{equation}

The anisotropic variational potentials are therefore
\begin{equation}
V_{\Pi_u}(r;a_\perp,a_\parallel)
=
 m_g+\frac{1}{2m_g}\left(\frac{2}{a_\perp^2}+\frac{1}{2a_\parallel^2}\right)
-\kappa I_C^{(\Pi)}(r;a_\perp,a_\parallel)
+\sigma I_\sigma^{(\Pi)}(r;a_\perp,a_\parallel)
-G_P\mathcal I_{\Pi}(a_\perp,a_\parallel),
\end{equation}
\begin{equation}
V_{\Sigma_u^-}(r;a_\perp,a_\parallel)
=
 m_g+\frac{1}{2m_g}\left(\frac{1}{a_\perp^2}+\frac{3}{2a_\parallel^2}\right)
-\kappa I_C^{(\Sigma)}(r;a_\perp,a_\parallel)
+\sigma I_\sigma^{(\Sigma)}(r;a_\perp,a_\parallel)
-G_\Sigma\mathcal I_{\Sigma}(a_\perp,a_\parallel).
\end{equation}
\end{widetext}
At fixed $r$, the adiabatic surfaces are obtained by minimizing with respect to $a_\perp$ and $a_\parallel$.
\begin{align}
&V_{\Pi_u}(r)=\min_{a_\perp,a_\parallel}V_{\Pi_u}(r;a_\perp,a_\parallel),
\nonumber\\
&V_{\Sigma_u^-}(r)=\min_{a_\perp,a_\parallel}V_{\Sigma_u^-}(r;a_\perp,a_\parallel).
\end{align}

\subsection{Connection to isotropic ansatz}
It is essential to verify that the spherical ansatz of Sec.~III is recovered in the isotropic limit. Setting
\begin{equation}
a_\perp=a_\parallel=\frac{1}{\beta},
\end{equation}
one finds
\begin{equation}
\psi_{\Pi}^{(m=\pm 1)}\propto \rho\,e^{im\phi}e^{-\beta^2(\rho^2+z^2)/2}
\propto r\,e^{-\beta^2 r^2/2}Y_{1,\pm 1}(\hat x),
\end{equation}
and
\begin{equation}
\psi_{\Sigma}\propto z\,e^{-\beta^2(\rho^2+z^2)/2}
\propto r\,e^{-\beta^2 r^2/2}Y_{10}(\hat x).
\end{equation}
The kinetic terms reduce correctly to
\begin{equation}
\langle \vec p^{\,2}\rangle_{\Pi}=\frac{5}{2}\beta^2,
\qquad
\langle \vec p^{\,2}\rangle_{\Sigma}=\frac{5}{2}\beta^2,
\end{equation}
as required for the isotropic $P$-wave multiplet. The instanton overlap also collapses to the Sec.~III result,
\begin{equation}
\mathcal I_{\Pi}=\mathcal I_{\Sigma}=\left(\frac{\beta^2}{\beta^2+\rho_0^{-2}}\right)^{5/2}.
\end{equation}
The Coulomb and confining matrix elements likewise reduce to the corresponding spherical $P$-wave integrals when $a_\perp=a_\parallel$.

The anisotropic Gaussian ansatz remains smooth at $r\to 0$, where one expects $a_\perp=a_\parallel$ by restoration of rotational invariance, but it also allows the variational solution at finite $r$ to develop an elongated longitudinal profile, $a_\parallel>a_\perp$, in response to the flat string-like valley along the molecular axis. It is therefore the minimal extension of the Gaussian midpoint ansatz consistent with the actual symmetry of the molecular problem.

\section{Simplified variational LFWF}
In this Appendix, we simplify the variational analysis to only the kinetic contribution to the Hamiltonian
\begin{equation}
U_{\rm kin}(x)
=
\sum_{i=1}^4 \frac{\omega_i}{x_i},
\qquad
\omega_i\equiv \langle M_i^2+p_{i\perp}^2\rangle ,
\label{eq:Ukin_omega_2}
\end{equation}
The light-front momentum fractions are restricted to the thetrahedral-cup
\begin{equation}
x_i\ge 0,
\qquad
\sum_{i=1}^4 x_i=1,
\qquad
x_g\equiv x_4.
\label{eq:xi_constraint_2}
\end{equation}
The kinetic contribution diverges at every face $x_i=0$ of the tetrahedron. Therefore the longitudinal WF must vanish there, exactly as in the baryonic triangular-cup problem~\cite{Shuryak:2022thi,Shuryak:2026pqt}.

For the ground state we can proceed variationally using the  Dirichlet ansatz
\begin{equation}
\varphi_4(x_1,x_2,x_3,x_g)
=
N^{(\parallel)}_4\,
(x_1x_2x_3)^{\alpha_q}\,x_g^{\alpha_g},
\label{eq:dirichlet4_only}
\end{equation}
with
\begin{equation}
\alpha_q>0,
\qquad
\alpha_g>0.
\label{eq:alpha_pos}
\end{equation}
and the normalization
\begin{equation}
\bigl|N_4^{(\parallel)}\bigr|^2
=
\frac{\Gamma(6\alpha_q+2\alpha_g+4)}
{\Gamma(2\alpha_q+1)^3\,\Gamma(2\alpha_g+1)}.
\label{eq:N4_dirichlet_2}
\end{equation}

For the  Dirichlet ansatz the expectation value of the longitudinal kinetic term is 
\begin{equation}
\langle U_{\rm kin}\rangle
=
3\omega_q\left\langle \frac{1}{x_1}\right\rangle
+\omega_g\left\langle \frac{1}{x_g}\right\rangle,
\label{eq:Uavg_start}
\end{equation}
where $\omega_1=\omega_2=\omega_3\equiv \omega_q$ and $\omega_4\equiv \omega_g$. Using (\ref{eq:dirichlet4_only}) we have
\begin{eqnarray}
&\left\langle \frac{1}{x_1}\right\rangle
=
\frac{6\alpha_q+2\alpha_g+3}{2\alpha_q},
\nonumber\\
&\left\langle \frac{1}{x_g}\right\rangle
=
\frac{6\alpha_q+2\alpha_g+3}{2\alpha_g},
\label{eq:inverse_moments}
\end{eqnarray}
so that
\begin{equation}
\langle U_{\rm kin}\rangle
=
\left(6\alpha_q+2\alpha_g+3\right)
\left(
\frac{3\omega_q}{2\alpha_q}
+\frac{\omega_g}{2\alpha_g}
\right).
\label{eq:Uavg_final}
\end{equation}
Minimization with respect to $\alpha_q$ and $\alpha_g$ gives
\begin{equation}
\frac{\alpha_g}{\alpha_q}=\sqrt{\frac{\omega_g}{\omega_q}},
\label{eq:ratio_alpha}
\end{equation}
and
\begin{equation}
\alpha_q=\frac{3+2r}{2(3+r)},
\qquad
\alpha_g=\frac{r(3+2r)}{2(3+r)},
\qquad
r\equiv \sqrt{\frac{\omega_g}{\omega_q}}.
\label{eq:alphas_solution}
\end{equation}
In the simplest constituent estimate, assuming
\begin{equation}
\omega_q\simeq m^2,
\qquad
\omega_g\simeq M_g^2,
\label{eq:omega_mass_only}
\end{equation}
with
\begin{equation}
m=0.35~{\rm GeV},
\qquad
M_g=0.90~{\rm GeV},
\label{eq:mass_values}
\end{equation}
we obtain
\begin{equation}
r=\frac{M_g}{m}\simeq 2.57,
\qquad
\alpha_q\simeq 0.731,
\qquad
\alpha_g\simeq 1.879.
\label{eq:alpha_num}
\end{equation}

\begin{figure}[h!]
    \centering
    \includegraphics[width=0.85\linewidth]{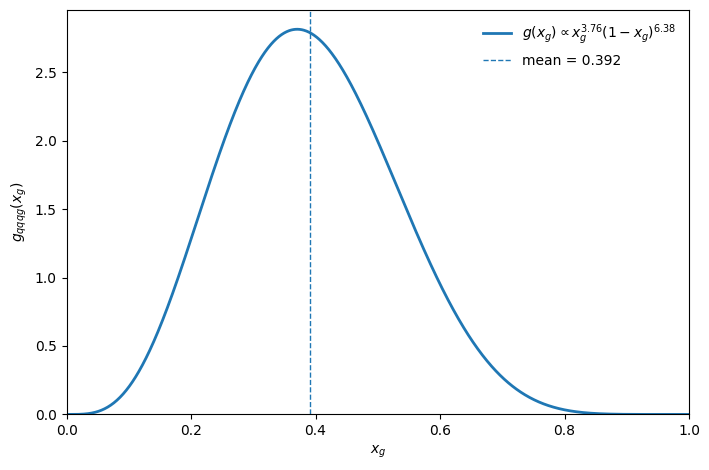}
    \caption{Gluon PDF for the $qqqg$ hybrid, including only the LF kinetic term or the tetrahedral-cup potential~\eqref{eq:xi_constraint_2}.}
    \label{fig_gluon_pdf_qqqg.pdf}
\end{figure}

The gluon PDF is
\begin{eqnarray}
&&g(x_g)
=\nonumber\\
&&\int_0^{1-x_g}dx_1\int_0^{\overline{x_g}=1-x_g-x_1}dx_2\,
\left|
\varphi_4(x_1,x_2,\overline{x_g},x_g)
\right|^2.\nonumber\\
\label{eq:gdef_dirichlet}
\end{eqnarray}
Carrying out the simplex integral explicitly gives
\begin{equation}
g(x_g)
=
\frac{\Gamma(6\alpha_q+2\alpha_g+4)}
{\Gamma(2\alpha_g+1)\Gamma(6\alpha_q+3)}
\,
x_g^{2\alpha_g}(1-x_g)^{6\alpha_q+2}.
\label{eq:gpdf_exact}
\end{equation}
For the variational values (\ref{eq:alpha_num}),
\begin{equation}
g(x_g)\propto x_g^{3.76}(1-x_g)^{6.38}.
\label{eq:gpdf_num}
\end{equation}
The small-$x_g$ power is controlled entirely by the gluon-face exponent $\alpha_g$ in the amplitude. The large-$x_g$ power receives two contributions: each of the three spectator quark factors contributes $(1-x_g)^{2\alpha_q}$ after squaring the wave function, and the two-dimensional spectator simplex measure provides the additional factor $(1-x_g)^2$.


\bibliography{main.bib}
\end{document}